# Atomic-scale visualization of the toroidal order in a trimeric Dy(III) single-molecule toroic

Michael J. Jenkins[1,#], Madalynn G. Marshall[2,5,#], Jie Xing[2,3,#], Muthu Satheeshkumar[4], Brandon D. Watson-Sanders[1], Xiaoping Wang[2], Christina M. Hoffmann[2], Gopalan Rajaraman[4,*], Huibo Cao[2,*], Rongying Jin[3,*], Zi-Ling Xue[1,*]

[1] *Department of Chemistry, University of Tennessee, Knoxville, Tennessee 37996, USA*

[2] *Neutron Scattering Division, Oak Ridge National Laboratory, Oak Ridge, Tennessee 37830, USA*

[3] *SmartState Center for Experimental Nanoscale Physics, Department of Physics and Astronomy, University of South Carolina, Columbia, South Carolina 29208, USA*

[4] *Department of Chemistry, Indian Institute of Technology Bombay, Mumbai 400076 Maharashtra, India*

[5] *Department of Chemistry and Biochemistry, Kennesaw State University, Kennesaw, Georgia 30144, USA*

[#] These authors contributed equally to this work. e-mail: rajaraman@chem.iitb.ac.in; caoh@ornl.gov; rjin@mailbox.sc.edu; xue@utk.edu

## Abstract

Single-molecule toroics (SMTs) offer a unique platform for next-generation quantum devices utilizing head-to-tail spin alignments in the compounds. Presence of toroidal moments in SMTs has been essentially based on magnetometry and *ab initio* calculations. Here, we report observation and probe of the toroidal moment in $[Dy_3(OH)(teaH_2)_3(paa)_3]Cl(OMe)$ [$teaH_3$: triethanolamine; paaH: *N*-(2-pyridyl)-acetoacetamide] from mapping of $Dy^{3+}$ magnetic susceptibility tensors by polarized neutron diffraction (PND). Neutron diffraction under variable magnetic fields demonstrates field-induced magnetization along the *c*-axis with toroidal moments antiparallelly stacked, providing definite proof of the field-induced order of toroidal moments. Temperature-dependence of neutron diffraction and magnetometry studies confirm the toroidal ground state below 16 K. For the first time, the combined use of PND, variable-field neutron diffraction, *ab initio* calculations, and magnetometry is introduced as a robust and quantitative methodology to probe molecular-scale toroidal magnetism. This integrated approach overcomes limitations of earlier indirect methods, establishes a benchmark framework for investigating SMTs, and provides valuable insights for the design of molecular quantum materials.

## Introduction

Single-molecule toroics (SMTs), characterized by ring-shaped arrangements of magnetic moments, have been actively studied as potential molecular quantum materials.[1-9] Their unique structural, magnetic, and electronic properties make them promising candidates for qubits in quantum computing.[5] These potential applications take advantage of tunable properties of spin systems in SMTs, where electronic and magnetic structures of these compounds can be influenced by coordination environments of magnetic ions.[6]

Toroidal moments are generated by non-collinear, head-to-tail alignments of magnetic moments (Fig. 1a,b) within, for example, a cluster molecule.[9] Toroidal spin states can be formed from a planar arrangement of spins or non-coplanar spins that couple in a circular pattern, leading to the cancellation of net magnetization within the toroidal plane, although there may still be out-of-plane dipole contributions to the overall spin states.[10,11] Toroidal moments are odd under either space inversion (*e.g.*, center inversion or mirror reflection) or time reversal, providing an effective method to gain spin chirality[12] and showing extraordinary potential for magnetoelectric coupling.[13] Toroidal moments can also align in the same direction, displaying ferrotoroidicity,[14,15] which is known as the fourth ferroic order, besides ferromagnetism, ferroelectricity, and ferroelasticity.[11,16]

Since the initial reports of SMTs,[1,2] the field has developed rapidly.[6,9-11,14,15,17-34] Most SMTs are lanthanide-based compounds, using large spins of the $Ln^{3+}$ ions such as $Dy^{3+}$. There have been reports of planar rings from $Dy_3$ up to $Dy_6$, mixed transition metal-lanthanide rings and clusters, as well as non-planar SMTs.[10,11,14,15,17-30,32,33,35-41] Other $Ln^{3+}$ ions such as $Ho^{3+}$ and $Tb^{3+}$ have also been reported with toroidal moments.[10,22,26,31,42] Typically toroidal moments of SMTs have been probed by magnetometry, including μSQUID,[1,7-9,14,15,19,27,30,38,43,44] where changes to net magnetic moments are observed at low temperatures, when the magnetic moments start to form their characteristic ring-shaped spin structures. These changes in magnetic moments can lead to non-magnetic ground states, as reported by Chibotaru, Ungur, and Soncini.[2] However, the loss of a net magnetic moment does not necessarily point to the presence of a toroidal moment in a cluster compound.[45] Additional measurements with other advanced techniques, such as cantilever torque magnetometry, high-field EPR, muon spin relaxation, far-IR, $^{161}$Dy-Mössbauer, and isotopically enrichments with $^{164}$Dy (with $I = 0$) and $^{163}$Dy (with $I = 5/2$), are used to examine ground states and magnetic excitations.[28,44-50] At magnetic excited states, there is no longer an absence of a magnetic moment, thus allowing for perturbation of spin states within the SMT compounds through the excitations.[2,46] These experiments, as well as associated theoretical studies and data fittings, have led to simulations of directions, intensities, and couplings of magnetic moments in SMTs. However, these techniques are indirect methods and cannot unambiguously determine the net toroidal moment due to the loss of local spin information in the crystal lattice.[51] In other words, the toroidal order in SMTs has not been experimentally validated at the atomic scale. In this work, we have investigated a $Dy_3$ SMT using the combination of polarized neutron diffraction (PND), variable-magnetic-field single-crystal neutron diffraction, and magnetometry to unambiguously determine the toroidal moment. This work provides for the first time the atomic-scale visualization of the toroidal order in an SMT.

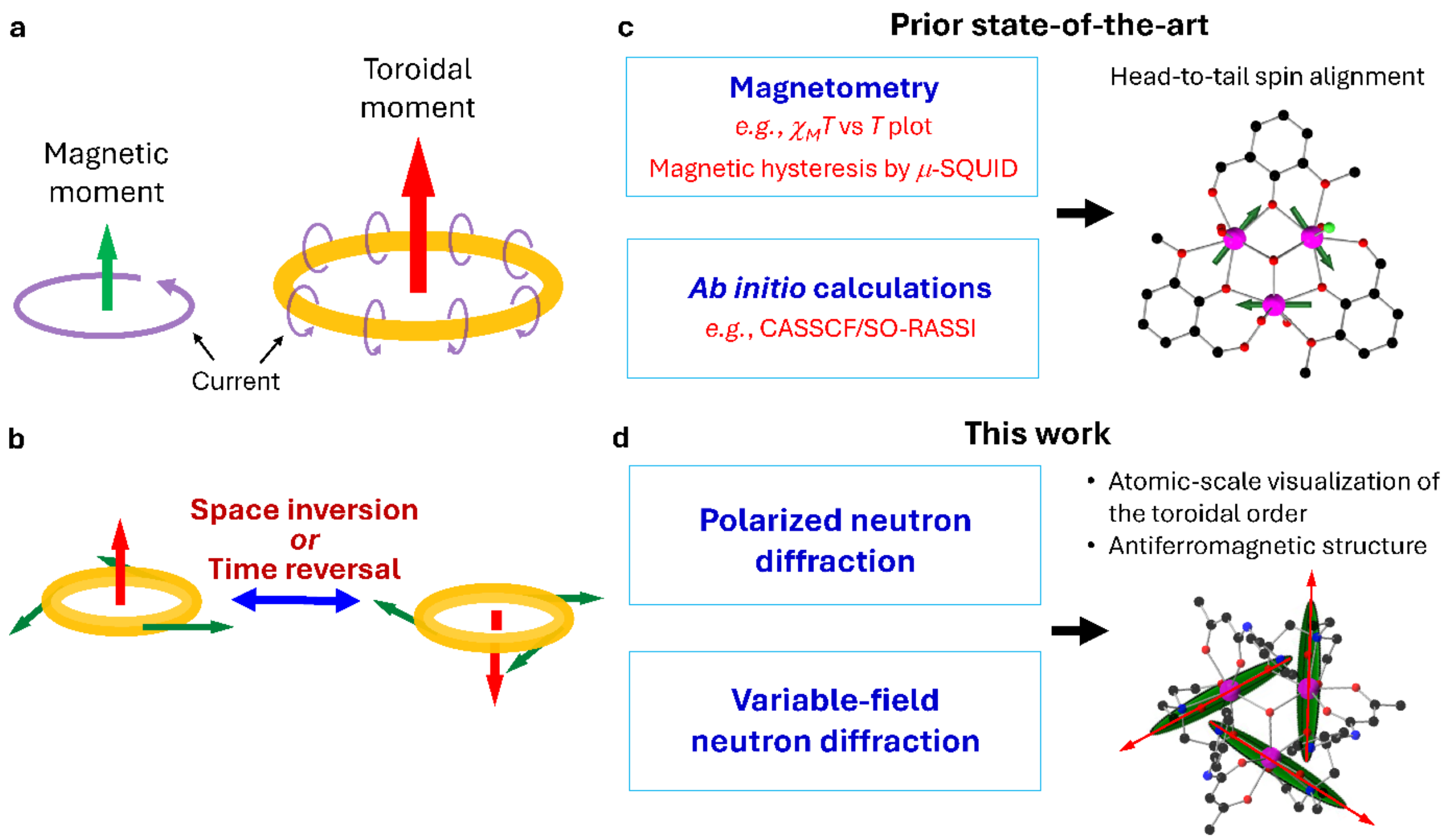

**Fig. 1 | Toroidal moment and methods for its detection in SMTs. a.** Current loops on a toroid surface producing a toroidal (magnetic) moment.[9] **b.** Toroidal moment from the vortex-like configuration of magnetic moments, such as those in three $Dy^{3+}$ ions in SMTs. The toroidal moment is odd under both space inversion and time reversal. That is, the toroidal moment flips or reverses its direction by the operations**. c-d.** Methods to determine toroidal moments in SMTs. $\chi_M$: molar magnetic susceptibility; $T$: temperature; $M$: magnetization; $H$: magnetic field; $\mu$-SQUID: micro-superconducting quantum interference device; CASSCF/SO-RASSI: Complete Active Space Self-Consistent Field/Spin-Orbit Restricted Active Space State Interaction.

Langley, Rajaraman, Murray and coworkers reported that $[Dy_3(OH)(teaH_2)_3(paa)_3]Cl_2 \cdot MeCN \cdot 4H_2O$ (**$Dy_3$-1**) showed SMT properties based on magnetometry studies of powder samples and *ab initio* calculations.[26] This compound was initially selected for the current studies because it crystallizes in trigonal space group $P\bar{3}$ (No. 147) with a $C_3$ axis through the center of the molecule, making the three $Dy^{3+}$ ions equivalent. Such high symmetry in the compound protects the triangular configuration of $Dy^{3+}{}_3$ moments and leads to a fully-magnetic-compensated, antiferromagnetic state within the triangular plane, which is determined to be a toroidal moment by PND and neutron diffraction measurements. As discussed below, we have instead prepared an analog of **$Dy_3$-1**, $[Dy_3(OH)(teaH_2)_3(paa)_3]Cl(OMe) \cdot 3MeOH \cdot PhCN$ (**$Dy_3$-2**), in a different solvent mixture to grow large, stable crystals in space group $P\bar{3}c1$ (No. 165).

It is challenging to observe toroidal moments experimentally, because metal ions, acting as electromagnetic particles with net magnetic dipoles or electric charges, can couple to the toroidal moment or obscure their signals. Therefore, highly sensitive techniques are required to observe toroidal moments. In the case of SMTs, determining susceptibility tensors of magnetic ions within the molecules is an efficient way to examine toroidal spin alignment, providing its definite proof. Experimental determination and probe of the head-to-tail spin alignment would allow for a better understanding of toroidal moments and their orders. To experimentally determine this toroidal moment, we have sought the combined use of PND, single-crystal neutron diffraction under applied magnetic fields, and magnetometry to determine the alignment of spins inside the molecule.

Magnetometry, often performed on powder samples, is routinely used to study molecular magnetic materials including SMTs, providing magnetization measurements to determine the overall magnetic behavior. Magnetic anisotropy of SMT single crystals has also been used to model the direction and magnitude of the atomic magnetic susceptibility and couplings between magnetic ions.[10] Ultra-sensitive $\mu$-SQUID magnetometry as a cornerstone technique has played a critical role in the studies of SMTs.[1,7-9,12,14,15,19,27,30,38]

Polarized neutron diffraction (PND) is a unique technique to probe molecular magnetism,[52-64] as reviewed by Luneau and Gillon[64] and one of us.[52] In PND, spin-up and spin-down neutrons are alternately generated and scattered by, e.g., a single crystal inside a small external magnetic field.[52] The magnetic field causes unpaired electrons in the sample to align "up" at a low temperature (e.g., 2 K). Spin-up and spin-down incident neutrons interact differently with the field-aligned "up" electron spins, leading to different diffraction intensities, even though the nuclear scattering from the atoms in the sample stays the same. Difference in the diffraction intensities is used to extract the magnetization density of unpaired electrons in the crystal, and the information can be further used to determine a local magnetic susceptibility tensor. In a PND experiment, Bragg peak intensities of spin-up and spin-down neutrons are measured to obtain flipping ratios of diffraction by the two types of neutrons, as described by Eq. 1:

$$R(k) = R(hkl) = \frac{I^{+}}{I^{-}} = \frac{{F_N}^2 + 2pq^2F_NF_M + q^2F_M^2}{{F_N}^2 - 2epq^2F_NF_M + q^2F_M^2} \tag{1}$$

where $I^{+}$ and $I^{-}$ = diffracted intensities at a measured Bragg position *k* by spin-up and spin-down incident neutrons, respectively; *q* = sine of the angle between the scattering vector and the magnetic field direction; *p* = polarization of the incident beam; *e* = flipping efficiency.

By fitting the flipping ratios, magnetic susceptibility tensors can be determined.[59,65]

PND has been effective in determining spin arrangements for both *d*- and *f*-block compounds.[52-54,56-61,63,66,67] Klahn and coworkers[60] studied magnetic anisotropy of a $Dy^{3+}$-based single-molecule magnet (SMM) by PND to determine the $Dy^{3+}$ magnetic susceptibility tensor. Similarly, PND has been used to compare similarities and differences of $Dy^{3+}$ and $Ho^{3+}$ SMMs $[Ln(H_2O)_5(HMPA)_2]I_3$ (HMPA = hexamethylphosphoramide), showing that the transverse elements of the atomic susceptibility are significantly different.[61] Temperature evolution of local magnetic anisotropy in a spin ice compound has also been studied by PND.[62,66] Spin density in $[Fe_8O_2(OH)_{12}(tacn)_6]Br_{4.3}(ClO_4)_{3.7}\cdot 6H_2O$ ($Fe_8$pcl; tacn = 1,4,7-triazacyclononane) was determined by PND, showing which of the eight $Fe^{3+}$ ions have up- or down-spins, leading to antiferromagnetic (AFM) interactions in the cluster.[63] To our knowledge, PND has not been used to probe an SMT.

Single-crystal µSQUID and cantilever torque magnetometry provide highly sensitive probes of magnetic anisotropy and, through angular-dependent measurements, can constrain the orientations of the local magnetic axes.[30,68,69] However, both techniques measure an integrated response of all magnetic centers and molecular orientations within the crystal, so that reconstruction of the individual anisotropy axes—and consequently assignment of a toroidal spin arrangement—generally remains non-unique. This limitation is particularly relevant for toroidal states, for which the head-to-tail arrangement of local moments produces little or no net magnetic dipole moment. In contrast, PND determines the local magnetic susceptibility tensor at crystallographically resolved Dy sites, directly linking each local anisotropy axis to the molecular structure. It therefore provides a site-resolved experimental determination of the head-to-tail arrangement underlying the toroidal moment, while variable-field neutron diffraction further establishes the relative arrangement of toroidal moments between molecules in the crystal.

In the current work, PND directly reveals the local magnetic anisotropy that leads to a head-to-tail spin alignment in a $Dy_3$-based SMT, demonstrating the first visualization of toroidal spin ordering from the magnetic susceptibility tensor in an SMT. Variable-field neutron diffraction reveals toroidal moments are antiparallelly stacked. Magnetometry provides couplings consistent with a toroidal ground state. The strategy integrating PND, variable-field neutron diffraction, *ab initio* calculations, and magnetometry overcomes limitations of prior indirect methods to validate experimentally molecular-scale toroidal magnetism, setting a new standard to probe SMTs.

## Results and discussion

### Selection of the compound and its structure determination

For PND experiments, high-quality data acquisition requires a large single crystal (>1 $mm^3$) with well-defined morphology and high symmetry. Large crystals with a high-symmetry space group are particularly advantageous, as they enhance the interpretability and resolution of magnetic susceptibility tensors. Following the published procedure,[26] we prepared the reported compound $[Dy_3(OH)(teaH_2)_3(paa)_3]Cl_2$ in mixed methanol-acetonitrile (MeOH-MeCN) solution, and grew its crystals of **$Dy_3$-1** by layering the solution with diethyl ether ($Et_2O$). We then attempted to grow larger crystals by using slow solvent vapor diffusion (Fig. S1), which yielded larger crystals. However, crystals of **$Dy_3$-1** rapidly degraded (in ~1 min) when exposed to air, likely due to solvent loss in the crystals. As a result, single-crystal neutron diffraction (SCND) of **$Dy_3$-1** at TOPAZ (Single-Crystal Diffractometer) at Oak Ridge National Laboratory (ORNL) was unsuccessful. The diffraction data indicated pronounced smearing of Bragg peaks consistent with decomposition during the experiment.

To resolve this issue, we have used a different solvent mixture (PhCN and MeOH) in the synthesis and growth of crystals. In addition, rather than layering the solution of the product with diethyl ether,[26] slow vapor diffusion of diethyl ether into the solution, as detailed in Supplementary information (SI, Fig. S1), led to the synthesis of $[Dy_3(OH)(teaH_2)_3(paa)_3]Cl(OMe)\cdot 3MeOH\cdot PhCN$ (**$Dy_3$-2**) as large yellow crystals (e.g., 1.1 x 1.0 x 1.5 $mm^3$). These crystals are semi-transparent in hexagonal shape, as shown by photos of the crystal used in the PND experiment (Fig. S2). No degradation of **$Dy_3$-2** crystals was observed during crystal selection (about 3-4 min) or subsequent use. The **$Dy_3$-2** crystals were then covered with a thin layer of fluorinated grease. All three crystals maintained structural integrity throughout PND and two neutron diffraction experiments (with and without magnetic fields).

Single-crystal X-ray diffraction (SCXRD) of **$Dy_3$-2** at 100(2) K shows the structure of $[Dy_3(OH)(teaH_2)_3(paa)_3]Cl(OMe)\cdot 3MeOH\cdot PhCN$ with a dication $[Dy_3(OH)(teaH_2)_3(paa)_3]^{2+}$ containing a bridging hydroxide ($\mu_3$-$OH^-$, O-H bond length = 0.84(5) Å) ligand, as in the structure of **$Dy_3$-1**.[26] While there are two $Cl^-$ anions in **$Dy_3$-1**, **$Dy_3$-2** has one $Cl^-$ and one methoxide $OMe^-$ anion which forms hydrogen bonds with the three MeOH solvent molecules in the lattice. Structural details are given in SI. An alternative structure of **$Dy_3$-2**, based on the single-crystal X-ray diffraction, is $[Dy_3(O)(teaH_2)_3(paa)_3]Cl\cdot 4MeOH\cdot PhCN$ with a monocation $[Dy_3(O)(teaH_2)_3(paa)_3]^+$ containing a bridging oxide ($\mu_3$-$O^{2-}$) ligand. Because X-ray diffraction is relatively insensitive to H atoms, we could not be certain whether there is a $\mu_3$-$OH^-$ ligand (and an $OMe^-$ anion hydrogen-bonded with three MeOH molecules) or a $\mu_3$-$O^{2-}$ ligand (and four MeOH molecules) in the structure of **$Dy_3$-2**. To confirm the structure, SCND of **$Dy_3$-2** at 100(2) K was conducted at TOPAZ, as hydrogen atoms can be located by neutron diffraction.[70] Indeed, neutron diffraction establishes a triangular $Dy_3$ core of **$Dy_3$-2** in space group $P\bar{3}c1$ (No. 165) capped by a $\mu_3$-OH bridge in the dication $[Dy_3(OH)(teaH_2)_3(paa)_3]^{2+}$ (Fig. 2b,c). The H8 atom on the $\mu_3$-OH ligand was located in the neutron difference map as a negative peak ("hole") of -7.4 fm $Å^{-3}$ at the bridging site (Fig. 2b,c) and refined to an O8–H8 distance of 0.937(10) Å, confirming the $\mu_3$-$OH^-$ rather than an $\mu_3$-$O^{2-}$ oxo bridging ligand. Additional support for the presence of the $\mu_3$-$OH^-$ ligand is given in Fig. S4. The lattice $MeO^-$/MeOH region is disordered. A MeOH component (C1S–O1S–H1S) shares the same O site with a $MeO^-$ component (C1SC–O1SC). Inclusion of $MeO^-$ anion at this site, together with $Cl^-$, provides overall charge balance and is consistent with all O–H atoms located in the neutron structure.

The O-H bond length of 0.937(10) Å in the $\mu_3$-$OH^-$ ligand by the neutron diffraction is more accurate than 0.84(5) Å by the X-ray diffraction. This difference arises because neutron diffraction determines atomic nuclei, whereas X-ray diffraction locates electron density. X-ray-derived X–H distances are therefore typically shorter. Here, structure of **$Dy_3$-2** in trigonal space group $P\bar{3}c1$ (No. 165) with $D_{3d}$ crystallographic point group symmetry is confirmed by both SCXRD and SCND.

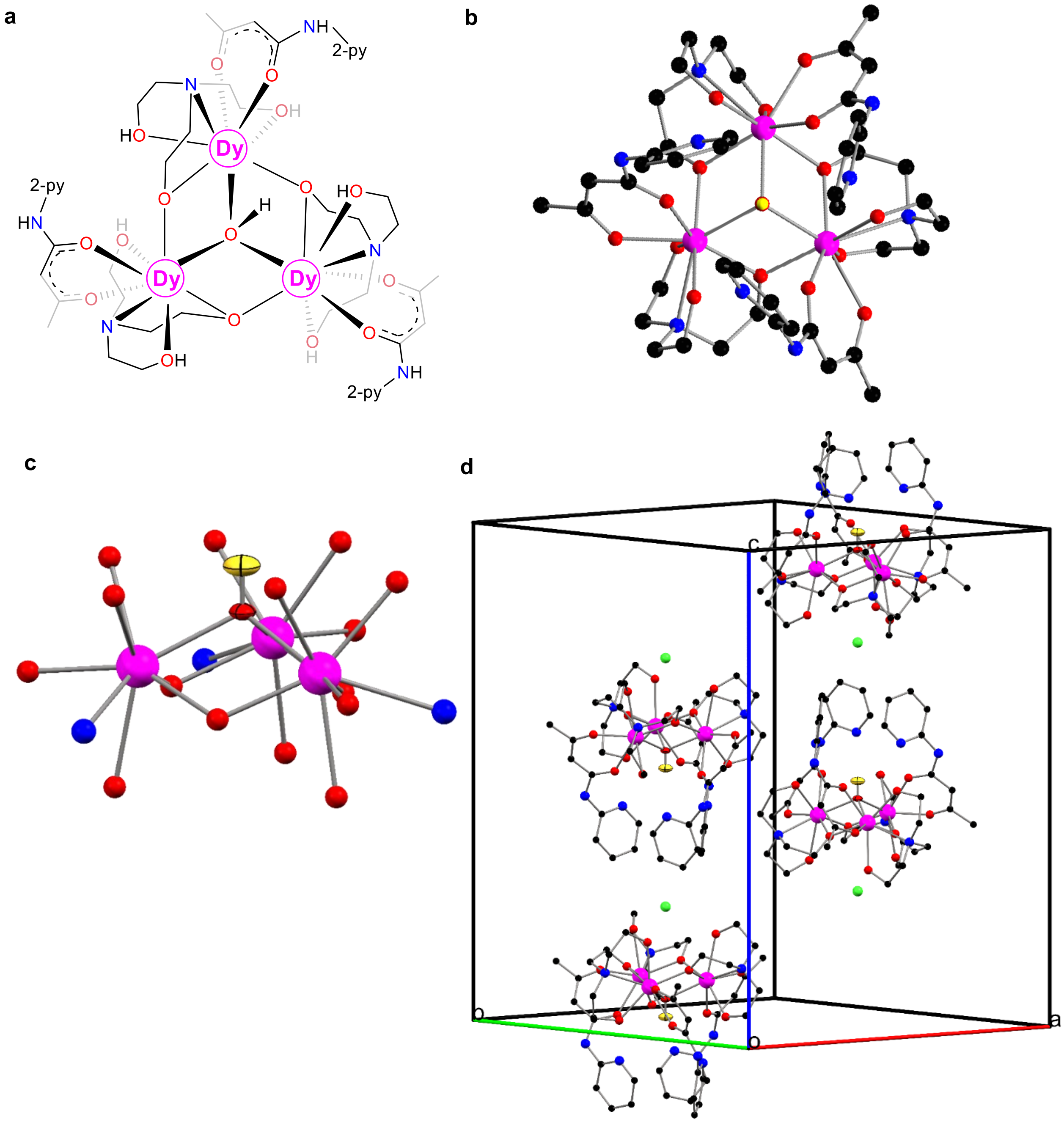

**Fig. 2 | Structure of $Dy_3$-2. a.** Chemical structure. **b.** Crystal structure of the $[Dy_3(OH)(teaH_2)_3(paa)_3]^{2+}$ cation in **$Dy_3$-2** by SCND. **c.** Structure of the $Dy_3$ core by SCND, showing thermal ellipsoids for the $\mu_3$-$OH^-$ ligand. **d.** Crystallographic unit cell by SCND containing four molecules in trigonal space group $P\bar{3}c1$ (No. 165) with $D_{3d}$ point group symmetry. $Dy^{3+}$ trimers show opposite stacking within the unit cell. In **b,d**, except the H atom in the $\mu_3$-$OH^-$ ligand, H atoms, $OMe^-$ anion, and solvent molecules are omitted for clarity.

**Magnetic properties**

The large crystals of **Dy3-2** have provided a unique opportunity to study the atomic magnetic susceptibility of **Dy3-2** using magnetometry and PND discussed below. For the polycrystalline sample (**$Dy_3$-$2_{poly}$**) between 2 and 300 K, magnetic susceptibility data were collected with an applied DC field of 0.1 T and isothermal magnetization at 1.9 K with the applied field varied from 0.0 to 7.0 T. These measurements were then also conducted with a single crystal oriented with the hexagonal face both parallel (**$Dy_3$-$2_{\parallel}$**) and perpendicular (**$Dy_3$-$2_{\perp}$**) to the magnetic field, as depicted in Fig. 3a. The data were plotted as $\chi$ vs $T$ (Fig. 3b), 1/$\chi$ vs $T$ (Fig. 3c), $M$ vs $H$ (Fig. 3d), and $\chi_M T$ vs $T$ (Fig. 3e) to allow for comparison. Note that the magnetic susceptibility measured at 0.1 T varies smoothly with temperature without any anomaly for both the single-crystal and polycrystalline forms. Interestingly, differences are observed when data is replotted as 1/$\chi$ vs $T$: it is approximately linear for **$Dy_3$-$2_{poly}$** but nonlinear for **$Dy_3$-$2_{\parallel}$** and **$Dy_3$-$2_{\perp}$**, suggesting anisotropic magnetic interactions.

We fit the magnetic susceptibility using the Curie-Weiss (CW) formula in Eq. 2 for both the polycrystalline sample and single crystal samples in the temperature range of 150–300 K:

$$\chi = \chi_0 + C/(T - \theta_{CW}) \qquad (2)$$

where $\chi_0$ = a constant representing the core contribution; $C$ = Curie constant; $\theta_{CW}$ = Curie-Weiss temperature.

The fits give $\chi_0$ = -0.6 × $10^{-3}$ emu/mol Oe for **$Dy_3$-$2_{\perp}$**, 1.3 × $10^{-3}$ emu/mol Oe for **$Dy_3$-$2_{\parallel}$**, and -0.9 × $10^{-3}$ emu/mol Oe for **$Dy_3$-$2_{poly}$**. The small and negative $\chi_0$ is typical for the core contribution. For the small and positive $\chi_0$ for **Dy3-$2_{\parallel}$**, it is likely dominated by the positive contribution from the sample holder. Nevertheless, the Curie-Weiss temperatures $\theta_{CW}$ = -151 K for **$Dy_3$-$2_{\perp}$**, 25.5 K for **$Dy_3$-$2_{\parallel}$**, and -15.8 K for **$Dy_3$-$2_{poly}$**. The different signs of the fitted Curie-Weiss temperatures for the two single-crystal orientations are consistent with strong anisotropy. Because crystal-field effects contribute substantially to these effective parameters, the negative and positive Curie-Weiss temperatures should not be assigned directly to microscopic FM or AFM exchanges. The fitted $C$ values correspond to the effective moment $\mu_{eff}$ values of 9.63 $\mu_B$ and 10.30 $\mu_B$ for **$Dy_3$-$2_{\perp}$** and **$Dy_3$-$2_{\parallel}$**, respectively. For **$Dy_3$-$2_{poly}$**, $\mu_{eff}$ ≈ 10.42 $\mu_B$, which is consistent with the moment of a $Dy^{3+}$ ion. In lanthanide systems, the Curie–Weiss temperature ($\theta_{CW}$) is strongly influenced by crystal-field splitting and magnetic anisotropy, and therefore does not directly reflect or quantify the underlying microscopic exchange interactions.[71,72]

We have measured the magnetic field dependence of the magnetization at various temperatures for **$Dy_3$-2$_\perp$**. As shown in Fig. 3d, *M* vs *H* is approximately linear at 30 K, 50 K, and 300 K. Deviation from linearity is clearly observed below 30 K, indicating increasing magnetic anisotropy and field-induced polarization.

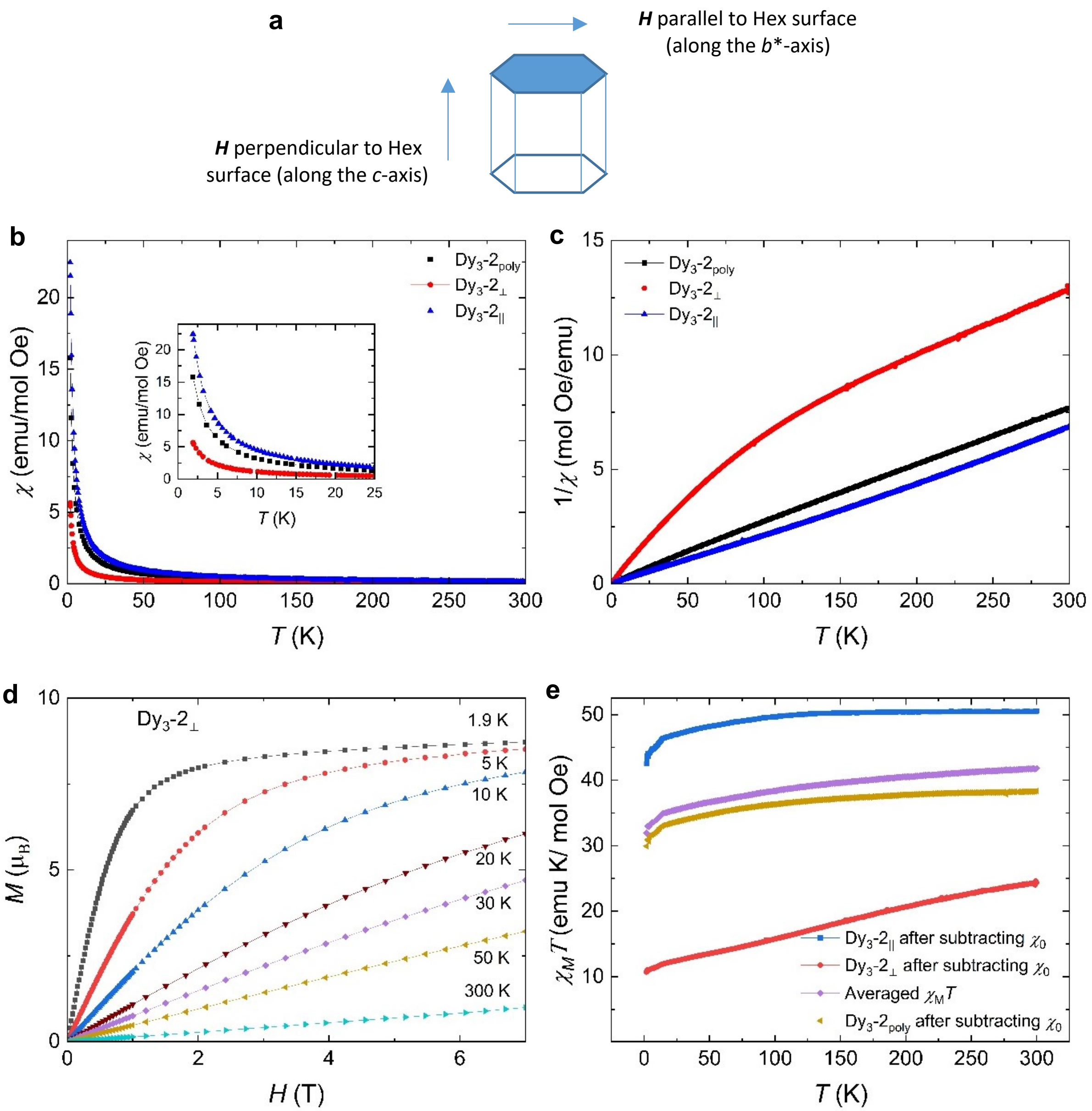

**Fig. 3 | Magnetic susceptibility data for powder and single-crystal samples of $Dy_3$-2. a.** Depiction of single-crystal orientation vs external field *H* for magnetic measurements. **b.** $\chi$ vs *T*. **c.** $1/\chi$ vs *T*. **d.** *M* vs *H* of **$Dy_3$-2$_\perp$** at indicated temperatures. **e.** $\chi_M T$ vs *T*. In the trigonal space group, the angle between *b*-axis and *b**-axis is 30° so that $b^* \perp a$ and *c*.

Since the estimated effective moment is nearly the same from both single-crystal and powder samples, we obtain the Dy contribution $\chi_M = \chi - \chi_0$. Fig. 3e shows the temperature dependence of $\chi_M T$ for **Dy₃-2**$_\perp$ and **Dy₃-2**$_\parallel$ and the polycrystalline sample. At 300 K, the $\chi_M T$ values are 50.6 and 24.4 $cm^3$ K $mol^{-1}$ for **Dy₃-2**$_\parallel$ and **Dy₃-2**$_\perp$, respectively. For comparison, we calculate the average $\chi_M T$ with the Poly_Aniso simulation $\chi_M T = 2/3(\chi_M T)_\parallel + 1/3(\chi_M T)_\perp$, which is presented in Fig. 3e as well. At 300 K, the average $\chi_M T$ = ~41.9 $cm^3$ K $mol^{-1}$, slightly higher than that for **Dy₃-2**$_{poly}$ (39.1 $cm^3$ K $mol^{-1}$). The average $\chi_M T$ from the single-crystal sample is consistent with the expected value of 42.5 $cm^3$ K $mol^{-1}$ for three non-interacting $Dy^{3+}$ ions, and is comparable to that of **Dy₃-1** which has a reported value of 41.63.[26]

In the $\chi_M T$ vs $T$ plot (Fig. 3e), there is an inflection point at 16 K and a broad maximum at 84 K for **Dy₃-2**$_\parallel$. For **Dy₃-2**$_\perp$, the corresponding features are less pronounced. At 2 K, the $\chi_M T$ values reach 42.4 and 10.7 $cm^3$ K $mol^{-1}$ for **Dy₃-2**$_\parallel$ and **Dy₃-2**$_\perp$, respectively, showing anisotropic magnetism at both low and room temperatures. The sharper decrease of $\chi_M T$ at 16 K for **Dy₃-2**$_\parallel$ is consistent with stronger in-plane magnetic anisotropy than for **Dy₃-2**$_\perp$. Comparison with the polycrystalline sample shows that single-crystal measurements resolve anisotropy that is obscured in the polycrystalline average.

Magnetization of all samples shows the characteristic field dependence of a $Dy_3$ SMT.[2,10] In the *M*-*H* plots in Fig. S7, a difference between increasing- and decreasing-field sweeps (i.e., hysteresis) is observed only for the powdered sample. This may be expected when examining the crystal structure of **Dy₃-2**, which appears to have an alternating stacking structure that could lead to the cancellation of net toroidal moments within the single crystal. The isothermal magnetization of a single crystal was collected with ***H*** parallel to the *a*, *b** and *c*-axes, respectively. Here, *a* and *b**-axes lie in the hexagonal plane and are 90° apart. The *c*-axis is perpendicular to the plane, giving **Dy₃-2**$_a$, **Dy₃-2**$_{b*}$, and **Dy₃-2**$_c$ (or **Dy₃-2**$_\perp$) data at 5 K in Fig. 4, respectively. Measurements were performed at 5 K to facilitate comparison with the PND experiment results discussed below. Significant magnetic anisotropy is shown in Fig. 4, suggesting that the coupling among the $Dy^{3+}$ ions within this molecule is retained regardless of direction of the external magnetic field. Interestingly, magnetization of the polycrystalline sample in the increasing field differs from that when the field is decreased (Fig. S7a). As *M* in the decreasing field condition is larger than that in the increasing condition, field-induced particle alignment may contribute to the powder sample measurement, indicating strong magnetic anisotropy. Nevertheless, magnetic anisotropy in single crystals can be detected by PND.

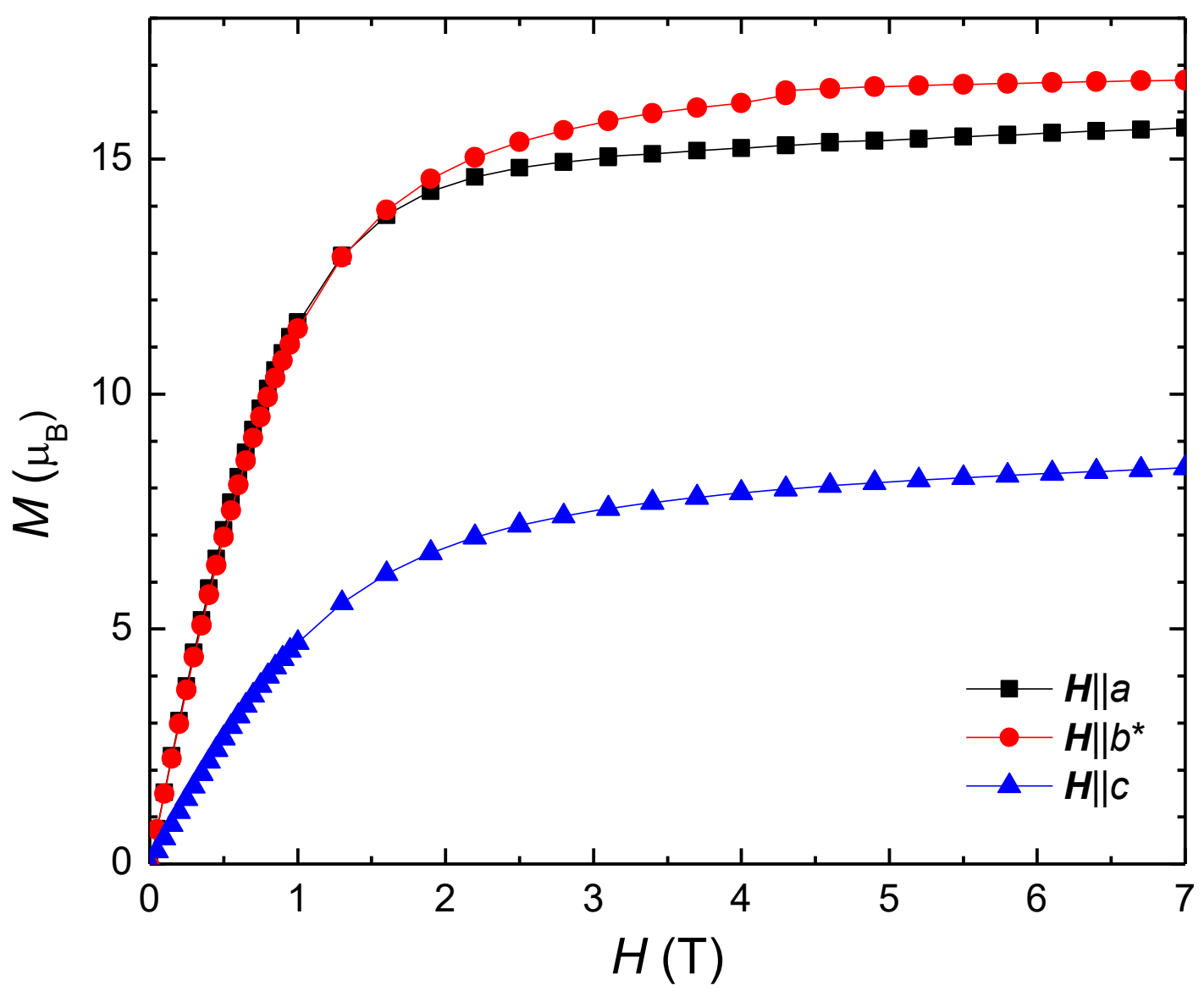

**Fig. 4 | Isothermal magnetization of the $Dy_3$-2 single crystal with the external magnetic field along *a*, *b**, and *c*-axes at 5 K.**

**Polarized neutron diffraction**

To understand the magnetic anisotropy of the $Dy^{3+}$ spins in **$Dy_3$-2**, we have measured the site-resolved magnetic susceptibility tensor of the $Dy^{3+}$ spin by PND. The reflections were measured in the spin-up and spin-down neutron channels with an applied magnetic field in the *ab*-plane where 27 good-quality flipping ratios were obtained (Eq. 1). The susceptibility tensor parameter $\chi_{33}$ were fixed to be consistent with the bulk magnetization measurements. Plot of experimental vs calculated flipping ratios in Fig. 5 shows a good refinement, giving the following refined magnetization tensor in Eq. 3 in unit of $\mu_B$/T:

$$\overleftrightarrow{\chi}_{Dy} = \begin{bmatrix} 1.3(2) & -1.5(3) & 1.0 \\ -1.5(3) & 3.5(3) & -1.0 \\ 1.0 & -1.0 & 1.8 \end{bmatrix} \qquad (3)$$

To compare with the magnetization in the bulk measurement, $\boldsymbol{M}_i$ ($i$ = *a*, $b^*$, and *c*, where the field is applied along) in Eq. 4 is used:

$$\boldsymbol{M}_i = \boldsymbol{H}_i \sum_j R_j \overleftrightarrow{\chi}_{Dy} \qquad (4)$$

$R_j$ = a crystallographic rotation operator that transforms the local site susceptibility tensor for the

three equivalent $Dy^{3+}$ sites.

Due to the crystallographic symmetry of the compound, this tensor is equivalent for all three $Dy^{3+}$ ions in the molecule, connected by a 3-fold rotation operator. The magnetization ellipsoids are shown in Fig. 5b, revealing that the overall spin arrangemet gives the closed head-to-tail arrangement characteristic of an SMT compound.

The nuclear structure of **$Dy_3$-2** determined by single-crystal neutron diffraction (SCND) at TOPAZ was used for the refinement of the flipping ratio data. However, since the SCND structure data and the PND data were collected at 100 K and 5 K, respectively, due to constraints of the instrument setup, the difference in measurement temperature may slightly affect the tensor determination slightly. Combining the PND-derived local anisotropy with the field-induced magnetic structure discussed below indicates antiparallel stacking of the toroidal magnetic moments under the applied field along *c* and cancellation of the net toroidal moment. This is because, among the four $Dy_3$ molecules in a unit cell (Fig. 2d), two trimers have the clockwise and two have the counterclockwise spin arrangement (Fig. 7), when the canted moments out-of-plane are polarized along the magnetic field direction ∥*c* (Fig. 7). Without the applied field, there is no order of the toroidal moment, consistent with a paratoroidic state.

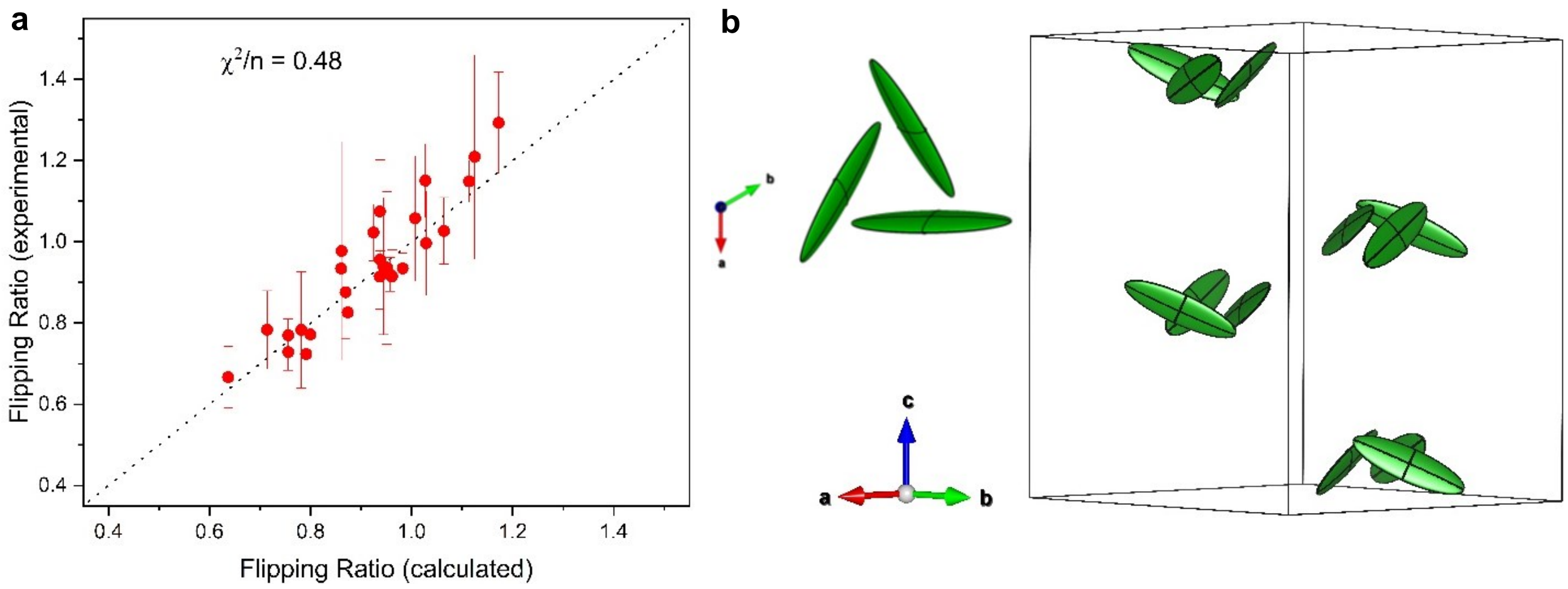

**Fig. 5 | PND results. a.** Observed and calculated flipping ratios from the local magnetic susceptibility refinement for **$Dy_3$-2**. **b.** The refined susceptibility tensor for **$Dy_3$-2** in the crystallographic unit cell. The head-to-tail spin alignment is indicated in the *ab*-plane for an antiferromagnetically coupled spin trimer, while the toroidal moment direction is field-controllable due to the slightly tilted ellipsoids to *c*-axis.

The susceptibility tensor for $Dy^{3+}$ ions relates the applied field *H* to the moment *M* at each $Dy^{3+}$ site by Eq. 4 and can therefore be used to compute the bulk magnetization for comparison with experimental data in Fig. 4. The total magnetization was obtained by summing the moments over all symmetry-related $Dy^{3+}$ sites and projecting the result along the field direction. Note, the susceptibility tensor indicates a local magnetic easy axis. Taking this information and the ordered moment in-plane of 8.0(5) $\mu_B$, determined by neutron diffraction at 4 T along *c*-axis (as discussed in the section below), the derived saturate magnetization in *ab*-plane at 2 K under a field ***H***‖*a* is $M_a$ = 16 $\mu_B$ per Dy-trimer. The value is comparable to the $M_a$ = 16.8 $\mu_B$ per Dy-trimer from the magnetization measurement (Fig. 4). Similarly, the small susceptibility component determined by PND (and the susceptibility tensor ellipsoid is nearly in *ab*-plane) is consistent with the smaller saturated magnetization, $M_c$ = 8.4 $\mu_B$ per Dy-trimer, when ***H***‖*c*. The PND and magnetization results are consistent with each other, showing the strong magnetic anisotropy of the $Dy_3^{3+}$ trimer. AFM coupled Dy-spins within the $Dy_3^{3+}$ trimer form a magnetic toroidal moment.

**Field-induced toroidal order by single-crystal neutron diffraction**

Single-crystal neutron diffraction has been used to examine the field-induced magnetic order and the toroidal order in solid-state materials. For **$Dy_3$-2**, two main features were examined: (1) Direction of the magnetic moment within each $Dy_3$ trimer; (2) Field-induced toroidal order due to the canted spins out-of-plane from antiparallel-stacked toroidal moments. Single-crystal neutron diffraction experiments were performed under magnetic fields applied parallel to the *c*-axis at 1.5 K. The magnetic field dependence of the Bragg peak (2 0 0) was measured between 0.0 T and 4.0 T in 0.1-T step. As shown in Fig. 6a, field-induced magnetic order along the *c*-axis emerges at ~0.2 T and reaches a maximum at ~2 T. Similar inflection points at ~2 T are also observed in the isothermal magnetization of **$Dy_3$-2** (Fig. 4) and other $Dy_3$ SMTs such as $[Dy_3(\mu_3\text{-}OH)_2L_3Cl_2(H_2O)_4][Dy_3(\mu_3\text{-}OH)_2L_3Cl(H_2O)_5]Cl_5\cdot 19H_2O$ and $[Dy_3(\mu_3\text{-}OH)_2L_3Cl(H_2O)_5]Cl_3\cdot 4H_2O\cdot 2MeOH\cdot 0.7MeCN$ (HL = ovanillin).[1] Temperature-dependent magnetic scattering at (2 0 0) was also measured between 1.5 K and 50.0 K with an applied magnetic field of 4.0 T. Below 10 K, the temperature was changed in 0.25-K steps to collect the magnetic scattering intensity. Above 10 K, the data were collected with 1-K steps. As shown in Fig. 6b, field-induced magnetic-ordering appears below 40 K with a slope-change around 16 K, consistent with the sharp change in the magnetic susceptibility data in Fig.3.

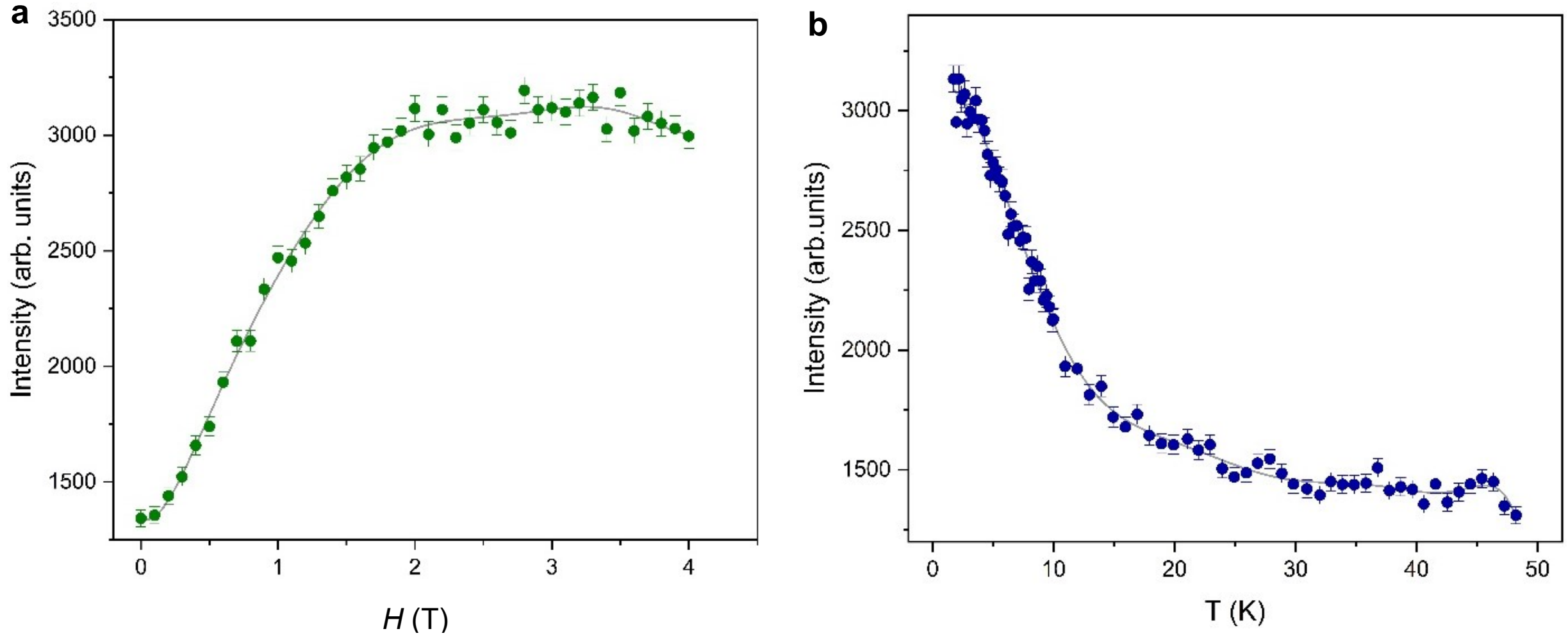

**Fig. 6 | Results of variable-field neutron diffraction at (2 0 0). a.** Field-dependent order parameter with data collected at 1.5 K with 0.1-T steps. **b.** Temperature-dependent order parameters for the magnetic form factor of **$Dy_3$-2** at 4.0 T. Field-induced magnetic order emerges at 0.2 T. Magnetic order appears below ~40 K with an applied magnetic field of 4.0 T.

Magnetic structure in **$Dy_3$-2** was derived by refining the single-crystal neutron diffraction data at 2.0 and 0.0 T. The scattering intensity at 0.0 T was subtracted from 2.0 T to obtain the magnetic scattering intensity. A magnetic symmetry analysis using Bilbao Crystallographic Server was carried out, considering *k* = 0 and the parent space group $P\bar{3}c1$ (No. 165).[73-75] The structure fits best with the magnetic space group $P\bar{3}c'1$ (No. 165.95).[76,77] For each of 230 crystallographic space groups (also known as Fedorov groups) which are for the symmetry of the nuclear structures of materials, orientations of unpaired electrons of atoms in the materials typically lead to a few magnetic space groups (also known as Shubnikov groups) for the parent space (Fedorov) group. The refined in-plane magnetic moment was 8.0(5) $\mu_B$, while the *c*-axis component was fixed at 2.25 $\mu_B$, based on the bulk magnetization data, to improve the refinement quality. The refinement yields two main results: (1) Antiferromagnetic order from antiparallel-stacked toroidal moments is observed along the *c*-axis, which is consistent with the bulk magnetic susceptibility measurements; (2) Head-to-tail orientation of the magnetic moments is within the *ab*-plane (Fig. 7), which agrees with the polarized neutron diffraction results. The orientation of spins within the *ab*-plane also shows the expected toroidal spin alignment for the magnetic $Dy^{3+}$ ions. A slight out of plane component of the magnetic structure is shown, consistent with the PND data collected. Note that the out-of-plane magnetic component allows an applied field along the *c*-axis to order the otherwise disordered toroidal moments. These results represent the first atomic-scale visualization of the toroidal order in an SMT.

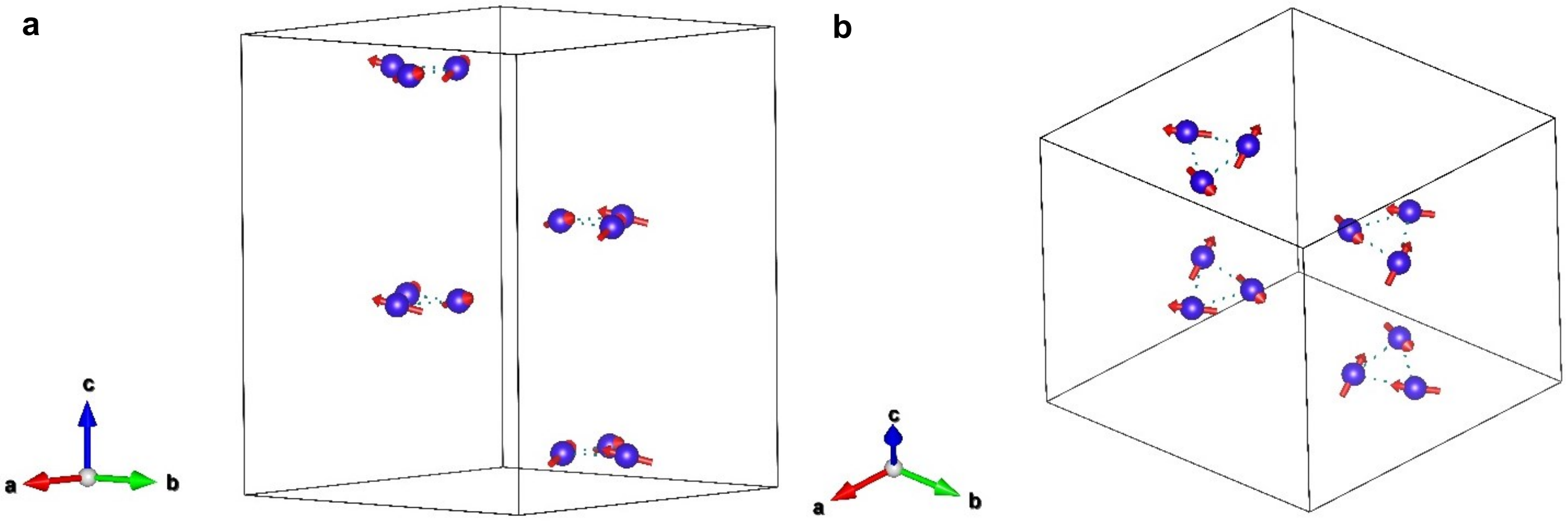

**Fig. 7 | Magnetic structure model of $Dy_3$-2 based on variable-magnetic-field neutron diffraction measurements and magnetic symmetry analysis using the Bilbao Crystallographic Server. a.** View in the *a*,*b*-plane; **b.** View along the diagonal line in the unit cell, respectively. There is a noticeable antiferromagnetic alignment of both toroidal moments and individual atomic magnetic moments.

**Theoretical analysis**

The magnetic properties were investigated using *ab initio* CASSCF/RASSI-SO calculations on model **1** with MOLCAS 8.0,[78] following previously reported methodologies[2,26] (details in SI). Model **1** is based on the crystal structure of **$Dy_3$-2** containing one $Cl^-$ anion, with solvent molecules and $OMe^-$ anion removed (Fig. 8a-b). Energies of Kramers doublets (KDs) from the ground $^6H_{15/2}$ multiplet of each $Dy^{3+}$ center are summarized in Table S4. The ground-state KDs at all three $Dy^{3+}$ sites exhibit strong Ising anisotropy, with a dominant $g_{zz}$ component close to the expected value for a nearly pure $M_J = \pm 15/2$ state (Table S4). The principal magnetic axes of the ground KDs are oriented with an out-of-plane angle $\theta \approx 17°$ and a canting angle $\phi \approx 40°$ (Table S4, Fig. 8a,b), consistent with the characteristic orientation associated with toroidal magnetic moments. (The canting angle represents the angular deviation of the magnetic axes relative to the tangential directions of an ideal circumference enclosing the triangular units.) The PND-derived canting angle ($\phi$) is approximately 35°, in agreement with calculated value. Since the spin-flip states do not alter either the canting angle or the out-of-plane angle, the comparison with the 5 K PND measurements remains valid. The calculated *g*-tensors and energy levels further reveal small transverse magnetic components for each $Dy^{3+}$ ion (Table

S4). Quadratic decomposition of the RASSI-SO wave functions for the eight lowest KDs projected onto the |$J$, $m_J$⟩ basis ($J$ = 15/2) shows that the ground KDs are dominated by the $m_J$ = ±15/2 contribution (~0.98), whereas the excited KDs exhibit substantial mixing and cannot be described in terms of pure $m_J$ states (Tables S5–S8). Overall, the calculated $g$-tensors, their orientations, and the wave-function decompositions are consistent with a low-symmetry ligand field at $Dy^{3+}$ sites in the **$Dy_3$–2** core.

To probe magnetic exchange interactions, Density Functional Theory (DFT) calculations were performed on model **1** using B3LYP/TZVP to obtain initial estimates of super-exchange coupling parameters ($J_{ex,\,\mathrm{Dy—Dy}}$ = –0.06 cm$^{-1}$; details in SI). This provides the exchange part of the magnetic coupling without the dipolar contribution. To extract the total exchange coupling, the Lines model[79] was employed, wherein the computed single-ion parameters, along with an initial guess for $J_{ex}$, were used as a single parameter to fit the susceptibility and magnetization data simultaneously. The best fit for the Lines model yields $J_{\mathrm{Ising,ex}} \approx 1.78$ cm$^{-1}$ for all interacting Dy–Dy pairs (Eq. S4 in SI). The dipolar contribution, calculated using Eq. S3, is about −0.2 cm$^{-1}$, resulting in a total interaction strength of $J_{\mathrm{Ising,tot}} \approx 1.6$ cm$^{-1}$ (Fig. 8c-e). The sign convention differs between the two exchange models: a positive $J_{\mathrm{Ising,ex}}$ corresponds to an antiferromagnetic $J_{\mathrm{Lines,ex}}$, reflecting the different definitions adopted in the Ising and Lines[79] formalisms.[44,80]

Theoretical evidence for the ground-state non-magnetic toroidal moment is provided by the direction-dependent Zeeman diagrams along the $x$, $y$ and $z$ directions in Fig. 8c, d and e, respectively. The toroidally ordered ground states are separated from the magnetic spin-flip states by an energy gap of 1.6 cm$^{-1}$. The ground-state magnetic moment of ~8.77 $\mu_B$ (Fig. S10) arises from uncompensated magnetic moments of the $Dy^{3+}$ centers within the circular toroidal arrangement and is markedly smaller than the value expected for three non-interacting $Dy^{3+}$ ions (~30 $\mu_B$). Although an ideal toroidal state would possess no net magnetic moment along the Z direction, the finite low-temperature magnetization observed here, consistent with previously reported triangular $Dy_3$ systems, indicates that model **1** does not behave as a conventional net toroidal SMT but instead displays mixed-moment single-molecule toroidal behavior. When the magnetic field is applied along the $z$ direction (Fig. 8c), the ground-state pseudo-doublet splitting increases with field owing to the residual uncompensated magnetic moment. In contrast, fields applied along the $x$ or $y$ directions (Fig. 8a,b) do not split the non-magnetic toroidal ground state; instead, the excited magnetic spin-flip states become energetically favored at fields of approximately 0.2 T.

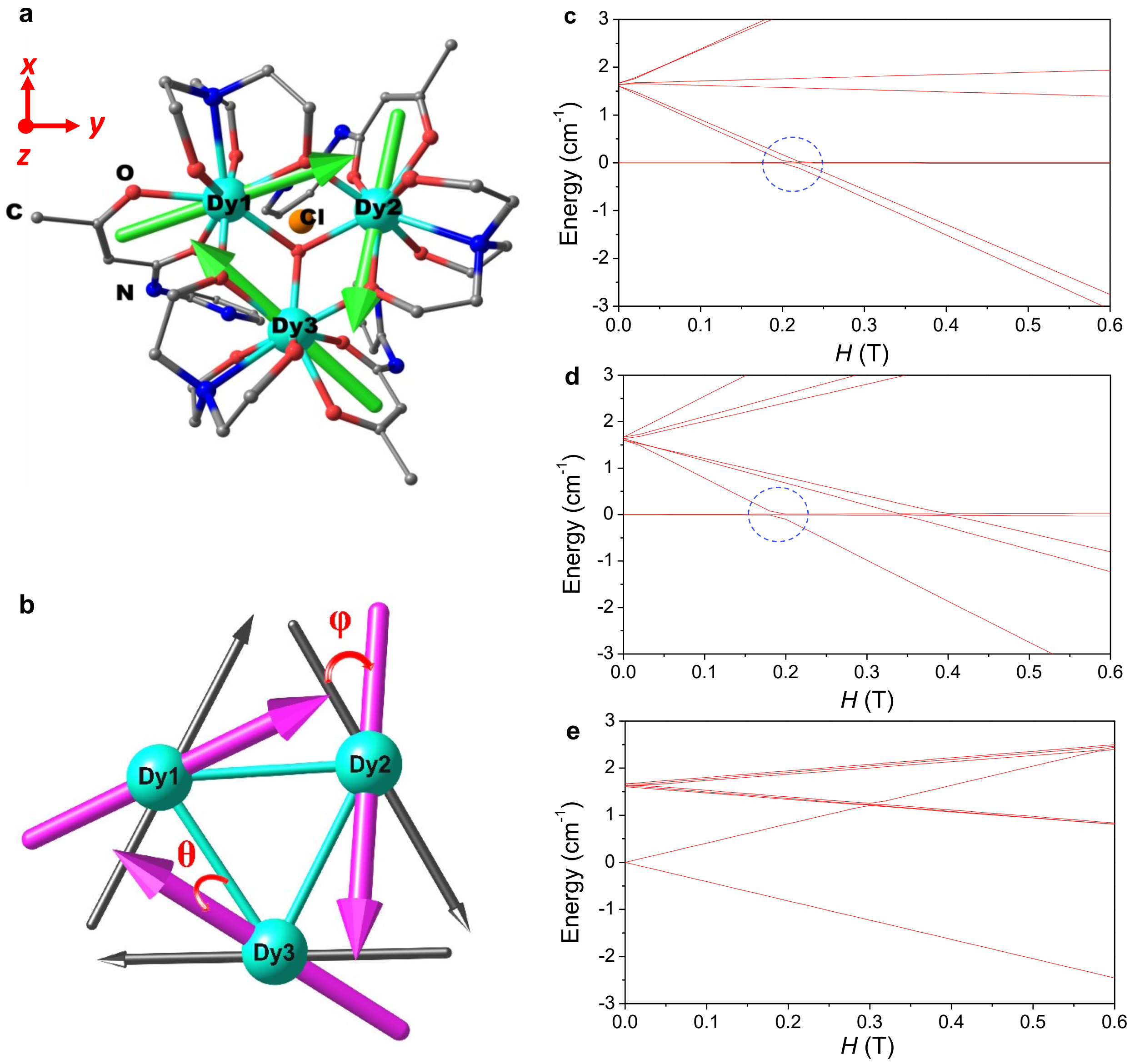

**Fig. 8. Results of *ab initio* calculations. a.** Local magnetic moments (green arrows) in the ground doublet in model **1**. H atoms are omitted for clarity. **b.** The $Dy_3$ core of **1** representing the canting angle (φ) and out-of-plane angle (θ). **c-e.** Zeeman splitting of **1** calculated using the POLY_ANISO lines model with magnetic fields applied along *x*, *y* and *z*, respectively. The dotted blue circles in **c** and **d** represent the magnetic states crossing the non-magnetic toroidal states.

Neutron diffraction uniquely enables accurate localization of hydrogen atoms, motivating an evaluation of the role of the $\mu_3$-OH proton in the **$Dy_3$–2** core. *Ab initio* calculations performed on model **2**, in which the $H^+$ ion is removed from the bridging $OH^-$ ligand in model **1**, giving the $[Dy_3(O)(teaH_2)_3(paa)_3]^+$ cation in model **2**, reveal a pronounced deviation of the magnetic anisotropy axes from the toroidal arrangement calculated for **1** with out-of-plane angle $\theta \approx 15.3^\circ$ and canting angle $\varphi \approx 70.6^\circ$ (Fig. S9). This substantial reorientation demonstrates that the proton in the bridging $OH^-$ ligand plays a critical role in stabilizing the toroidal state, suggesting that protonation state, and thus pH, may provide a viable handle for tuning toroidal behavior in SMTs.

In summary, we have used polarized neutron diffraction to unambiguously determine the head-to-tail spin alignment in $[Dy_3(OH)(teaH_2)_3(paa)_3]Cl(OMe)\cdot 3MeOH\cdot PhCN$ (**Dy$_3$-2**), mapping the atomic susceptibility tensor of $Dy^{3+}$ ions and confirming the toroidal moment in the SMT. Variable-field neutron diffraction has validated the expected toroidic order, showing the antiparallel stacking of the toroidal moments. **Dy$_3$-2** shows magnetic behavior consistent with an SMT in DC susceptibility studies, both for polycrystalline and single crystal samples. Results of *ab initio* CASSCF calculations are consistent with those from the experiments. This work is, to our knowledge, the first atomic-scale visualization of the toroidal order in an SMT.

## Methods

The following are given in SI: Synthesis of **Dy$_3$-2**, schematic of the setup for the growth of **Dy$_3$-2** crystals, and the structure determination by single-crystal X-ray diffraction.

### Single-crystal neutron diffraction to determine the structure of Dy$_3$-2

The TOPAZ diffractometer at ORNL was used to collect data for **Dy$_3$-2**.[81] A single crystal measuring 0.45 x 0.50 x 0.65 $mm^3$ in size was mounted on a Kapton pin and measured at 100(2) K under a nitrogen cryostream. Additional details for SCND are given in SI.

### Magnetic susceptibility measurements

DC susceptibility measurements were completed using a Quantum Design MPMS-7 T. Magnetization with variable temperatures and variable magnetic fields was collected. Measurements were taken of a powdered sample as well as single crystals oriented in multiple directions. Single crystals were coated with a thin layer of fluorinated grease to protect the crystal from degradation.

### Polarized neutron diffraction (PND)

The DEMAND diffractometer[82] at ORNL was used to perform single-crystal PND measurements with a polarized neutron beam of 1.542 Å from a bent Si-220 monochromator[83] and a calibrated neutron polarization of 76%. The sample was measured at 5 K with ***H***∥***ab*** and a permanent magnetic set was used to fix the field to 0.6 T. CrysPy software[59] was used to analyze measured flipping ratios.

### Neutron diffraction under magnetic fields

DEMAND[82] was also used to perform the field studies. The sample was measured at 1.5 K with variable magnetic fields from 0 to 4 T. Data subtraction and integration were performed using FulProf.[84,85]

### Data availability and associate content

Additional data and figures are provided in Supplementary information, including synthesis, crystallization, single-crystal X-ray diffraction, and computational details. Other data are available at figshare (https://doi.org/10.6084/m9.figshare.32338896).

### Acknowledgement

The authors thank Drs. Alexandria N. Bone, Pagnareach Tin, and Adam T. Hand for their help in the selection of the $Dy_3$ compound for the current study.

### Funding

US National Science Foundation supported this research at the University of Tennessee [CHE-2349345 to Z.L.X. and DGE-2152159 for an NRT (Research Traineeship Program) fellowship to B.D.W.-S.]. M.M. and H.C. acknowledge support by the U.S. Department of Energy (DOE), Office of Science, Office of Basic Energy Sciences, Early Career Research Program Award KC0402020, under Contract DE-AC05-00OR22725. The work at Kennesaw State University was supported by the U.S. Department of Energy, Office of Science, Office of Basic Energy Sciences under Award Number DE-SC0025735 to M.M. M.S. acknowledges the Prime Minister Research Fellowship (PMRF) provided by the Ministry of Education, Government of India. G.R. would like to thank DST/ANRF for funding (CRG/2022/001697; National Quantum Mission DST/QTC/NQM/QMD/2024/4). This research used resources at the High Flux Isotope Reactor (HFIR) and Spallation Neutron Source (SNS), DOE Office of Science User Facilities operated by the Oak Ridge National Laboratory. Beam times at DEMAND (Dimensional Extreme Magnetic Neutron Diffractometer, HB-3A) were allocated under proposals IPTS-29385, IPTS-30682, and IPTS-31432. Beam times at TOPAZ (Single-Crystal Diffractometer, BL-12) were allocated under proposals IPTS-28290 and IPTS-36415.

## Author contributions

M.J.J. synthesized and performed preliminary characterization of **$Dy_3$-1** and **$Dy_3$-2**, studied crystal stability, and developed the method to grow large crystals of **$Dy_3$-2**. M.J.J. and Z.-L.X. selected **$Dy_3$-1** for the initial studies. M.J.J., X.W., and C.M.H. solved the structure of **$Dy_3$-2** using single-crystal X-ray diffraction. B.D.W.-S., X.W., and C.M.H. solved the structure of **$Dy_3$-2** using single-crystal neutron diffraction on TOPAZ at ORNL. X.W. and C.M.H also conducted the single-crystal neutron diffraction of **$Dy_3$-1**. J.X. and R.J. performed magnetic susceptibility measurements and analyzed the results. M.M., M.J.J., H.C., and Z.-L.X. participated in the two neutron diffraction experiments on DEMAND at ORNL. M.M. completed neutron diffraction data analyses from the two experiments. M.S. and G.R. conducted *ab initio* calculations. Z.-L.X., H.C., and R.J. initiated the project. M.J.J. and Z.-L.X. wrote the manuscript with input from other authors.

## Competing interests

The authors declare no competing financial interest.

## Materials and correspondence

G. Rajaraman for *ab initio* calculations

H. Cao for PND and variable-field and variable-temperature neutron diffraction

R. Jin for magnetometry studies

Z-L. Xue for synthesis, crystal structures and other characterizations

## Supplementary information

# Atomic-scale visualization of the toroidal order in a trimeric Dy(III) single-molecule toroic

## Table of contents

## S1. Additional methods

The reactions were carried out in room temperature aerobic conditions. Unless otherwise noted, chemicals and solvents were obtained from commercial sources and used without purification. Triethanolamine (teaH$_3$, Fig. S1) was distilled before use. *N*-(2-pyridyl)-acetoacetamide (paaH, Fig. S1) was synthesized as previously reported, and purified using a silica gel column.[S1] Purity was confirmed by $^1$H NMR.

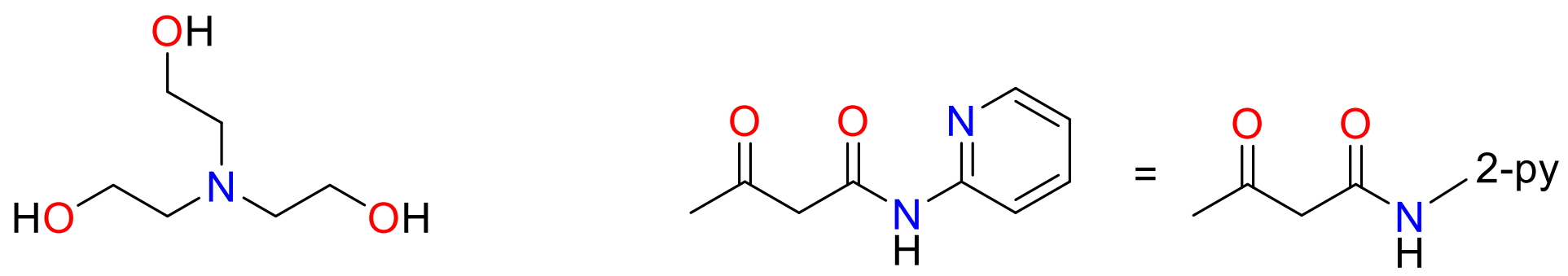

triethanolamine (teaH$_3$) *N*-(2-pyridyl)-acetoacetamide (paaH)

**Fig. S1 | Precursors to the organic ligands in [Dy$_3$(OH)(teaH$_2$)$_3$(paa)$_3$]Cl(OMe)·PhCN·3MeOH (Dy$_3$-2).**

### S1.1. Synthesis and crystallization of Dy$_3$-2

**Dy$_3$-2** was synthesized using a procedure similar to the reported.[S2] Solvents were changed from MeCN/MeOH to PhCN/MeOH. In addition, diethyl ether vapor diffusion (Fig. S2) was used to grow crystals large enough for single-crystal X-ray and neutron diffraction. Into 14 mL of a solution of PhCN/MeOH (1:1 v/v), 0.38 g (1 mmol) DyCl$_3$·6H$_2$O is dissolved (10 min mixing) followed by the addition of 0.18 g (1 mmol) paaH and 0.55 mL (4 mmol) triethylamine. The solution is mixed for a few minutes then 0.13 mL (1 mmol) triethanolamine (teaH$_3$) is added and the solution is mixed continuously for 3 hours. This mother liquor is then filtered through a fine filter and an ether vapor solution is set up. Crystals are initially seen within 3-4 days, with large crystals taking up to 4 weeks to form. Recorded yield of 43%. From this synthesis, hexagonal block-like crystals were grown. Neutron diffraction showed that the hexagonal crystals have significantly better signal intensities and thus were used.

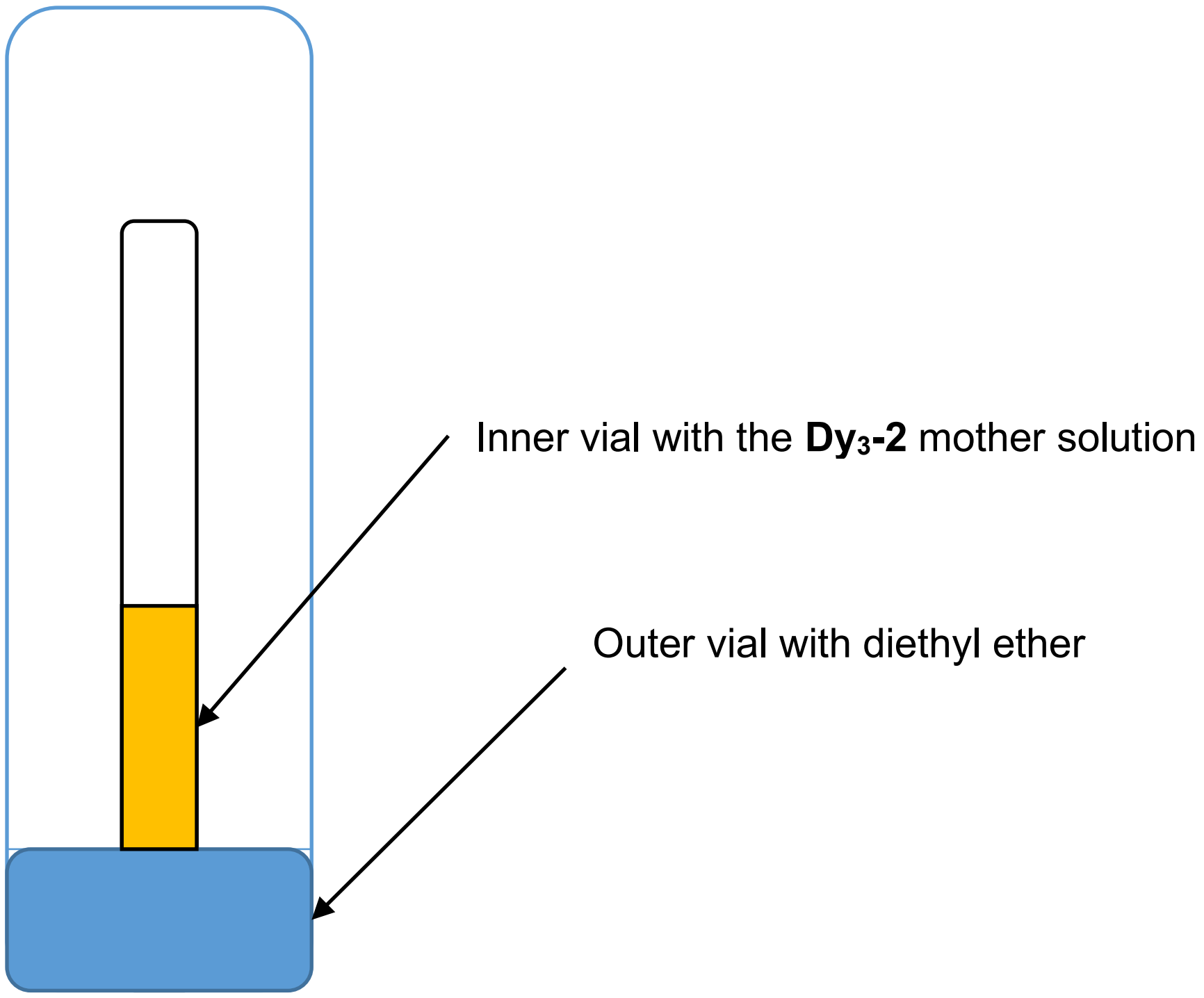

**Fig. S2 | Schematic of a diethyl ether vapor diffusion setup used for the growth of $Dy_3$-2 crystals.** The system is sealed, and the ether vapor slowly mixes with the mother solution of PhCN/MeOH to precipitate out crystals. Crystals are then removed from the walls of the inner vial for collection.

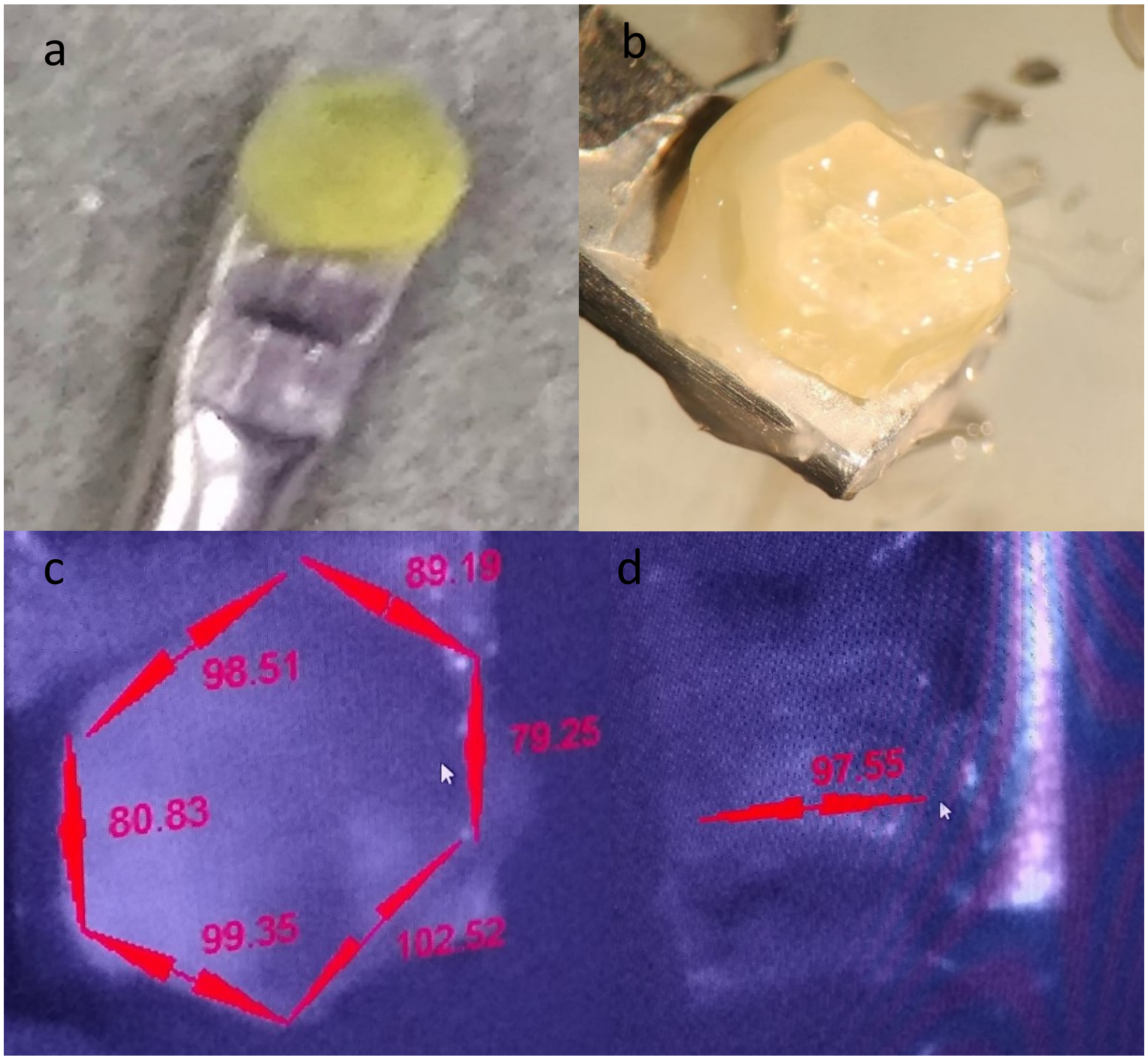

**Fig. S3 | Images of the crystal used in the PND experiment. a**, Sample before applying grease and measuring. **b**, Sample weeks after the experiment; **c,d**, Size of the crystal when mounted in the instrument (unit: 10 μm; "98.51" = 985.1 μm = 0.9851 mm).

### S1.2. Single-crystal X-ray diffraction

Single-crystal X-ray diffraction data for **$Dy_3$-2** was collected using a Bruker D8 Venture at 100 K with Mo Kα radiation. Data collection and integration was completed using APEX 3 programs, reduced using Bruker SAINT, and corrected for absorption using the SADABS multi-scan program. Reflections were merged using SHELXL, with computing graphics by Olex2. [S3-5]

### S1.3. Single-crystal neutron diffraction (SCND) at TOPAZ

The data-collection strategy was planned with *CrystalPlan*,[S6] and Bragg peaks were integrated using Mantid python program developed at TOPAZ.[S7] Standard TOF corrections (Lorentz factor, incident spectrum, detector efficiency) were applied during reduction.[7] The reduced data were saved in *SHELX HKLF2* format,[S8] with the neutron wavelength recorded separately for each reflection. The neutron crystal structure was refined with *SHELXL-2018/3*.[S8] The neutron refinement converged at $R_1$ = 0.0522 and $wR_2$ = 0.105.

## S2. Additional results and discussion

### S2.1. Crystal structure of $Dy_3$-2

Crystallographic data for **$Dy_3$-2** from SCND and SCXRD are compared with those of reported **$Dy_3$-1**[S9] in Table S1. Selected bond lengths (Å) and angles (°) in **$Dy_3$-2** from SCND and SCXRD are given in Table S2.

**Table S1**. Crystallographic data for **$Dy_3$-2** from SCND and SCXRD as well as **$Dy_3$-1** (reported by Langley et al.[S9])

| | **$Dy_3$-2 (neutron)** | **$Dy_3$-2 (X-ray)** | **$Dy_3$-1**[S9] |
|---|---|---|---|
| Formula | $C_{56}H_{90}ClDy_3N_{10}O_{20}$ | $C_{56}H_{90}ClDy_3N_{10}O_{20}$ | $C_{47}H_{81}Cl_2Dy_3N_{10}O_{20}$ |
| Formula weight | 1746.32 | 1746.32 | 1664.55 |
| Crystal size ($mm^3$) | 0.45 × 0.50 × 0.65 | 0.272 × 0.228 × 0.185 | 0.4 x 0.2 x 0.1 |
| Crystal System | Trigonal | Trigonal | Trigonal |
| Space group | $P\bar{3}c1$ (No. 165) | $P\bar{3}c1$ (No. 165) | $P\bar{3}$ (No. 147) |
| $a$ (Å)<br>$b$ (Å)<br>$c$ (Å) | 17.5217(7)<br>17.5217(7)<br>26.4332(15) | 17.5420(6)<br>17.5420(6)<br>26.4299(13) | 17.5986(3)<br>17.5986(3)<br>13.1888(5) |
| $\alpha$ (°)<br>$\beta$ (°)<br>$\gamma$ (°) | 90<br>90<br>120 | 90<br>90<br>120 | 90<br>90<br>120 |
| $V$ (Å$^3$) | 7028.0(7) | 7043.43 | 3537.46(16) |
| Wavelength (Å) | 0.4–3.5 (time-of-flight) | 0.71073 | 0.71073 |
| $T$ (K) | 100(2) | 100(2) | 123(2) |
| $Z$ | 4 | 4 | 2 |
| $\rho_{calc}$ | 1.650 | 1.647 | 1.555 |
| Data measured | 2944 | 269480 | 36880 |
| Independent reflections | 2596 | 7084 | 5400 |
| Parameters | 527 | 397 | 308 |
| Restraints | 108 | 73 | 51 |
| $R_1$ [$I > 2\sigma(I)$]<br>$wR_2$ | 0.0522<br>0.105 | 0.0227<br>0.0594 | 0.0247<br>0.0680 |
| Goodness of fit | 1.056 | 1.120 | 1.128 |

**Table S2.** Selected bond lengths (Å) and angles (°) in **$Dy_3$-2** from SCND and SCXRD

| **Bond Lengths** | | |
|---|---|---|
| | Neutron diffraction | X-ray diffraction |
| Dy1-Dy1 | 3.771(2) | 3.7792(2) |
| Dy1-O3 | 2.388(3) | 2.3897(16) |
| Dy1-O4 | 2.254(4) | 2.2520(16) |
| | 2.285(4) | 2.2891(16) |
| Dy1-O5 | 2.414(3) | 2.4133(15) |
| Dy1-O6 | 2.286(4) | 2.2821(17) |
| Dy1-O7 | 2.489(3) | 2.4884(16) |
| Dy1-O8 | 2.383(2) | 2.3833(9) |
| Dy1-N11 | 2.614(3) | 2.6134(19) |
| O8-H8 | 0.937(10) | 0.84(5) |
| O3S-H3S | 0.973(11) | 0.886(10) |
| **Bond Angles (°)** | | |
| | Neutron diffraction | X-ray diffraction |
| Dy1-O4-Dy1 | 112.34(15) | 112.65(6) |
| Dy1-O8-Dy1 | 104.62(12) | 104.91(6) |
| O4-Dy1-O3 | 153.86(12) | 154.00(6) |
| | 89.03(13) | 88.95(6) |
| O4-Dy1-O4 | 103.0(2) | 102.84(9) |
| O4-Dy1-O5 | 83.47(13) | 83.75(6) |
| | 141.32(12) | 141.32(5) |
| O4-Dy1-O6 | 90.47(15) | 90.47(6) |
| | 143.33(13) | 143.37(6) |
| O4-Dy1-O7 | 73.67(13) | 73.84(6) |
| | 77.06(13) | 77.03(6) |
| O4-Dy1-O8 | 71.78(12) | 71.53(5) |
| | 71.26(11) | 70.91(5) |
| O3-Dy1-O5 | 73.00(12) | 72.94(5) |
| O3-Dy1-O6 | 93.46(13) | 93.63(6) |
| O3-Dy1-O7 | 132.17(14) | 131.90(6) |
| O3-Dy1-O8 | 90.87(12) | 91.19(6) |
| O5-Dy1-O6 | 73.30(12) | 73.26(6) |
| O5-Dy1-O7 | 139.85(14) | 139.99(6) |
| O5-Dy1-O8 | 75.00(12) | 75.50(6) |
| O6-Dy1-O7 | 74.32(13) | 74.21(6) |
| O6-Dy1-O8 | 145.13(12) | 145.38(5) |
| O7-Dy1-O8 | 125.41(15) | 125.11(7) |
| O3-Dy1-N11 | 67.11(10) | 67.00(6) |
| O4-Dy1-N11 | 138.85(10) | 138.84(6) |
| | 68.71(11) | 68.94(6) |
| O5-Dy1-N11 | 129.02(11) | 128.68(6) |
| O6-Dy1-N11 | 78.66(12) | 78.54(6) |
| O7-Dy1-N11 | 65.18(11) | 65.00(6) |
| O8-Dy1-N11 | 134.06(7) | 134.11(6) |
| N11-Dy1-Dy1 | 153.05(6) | 153.04(4) |
| | 100.05(7) | 100.10(5) |
| Dy1-Dy1-Dy1 | 60.0 | 60.0 |

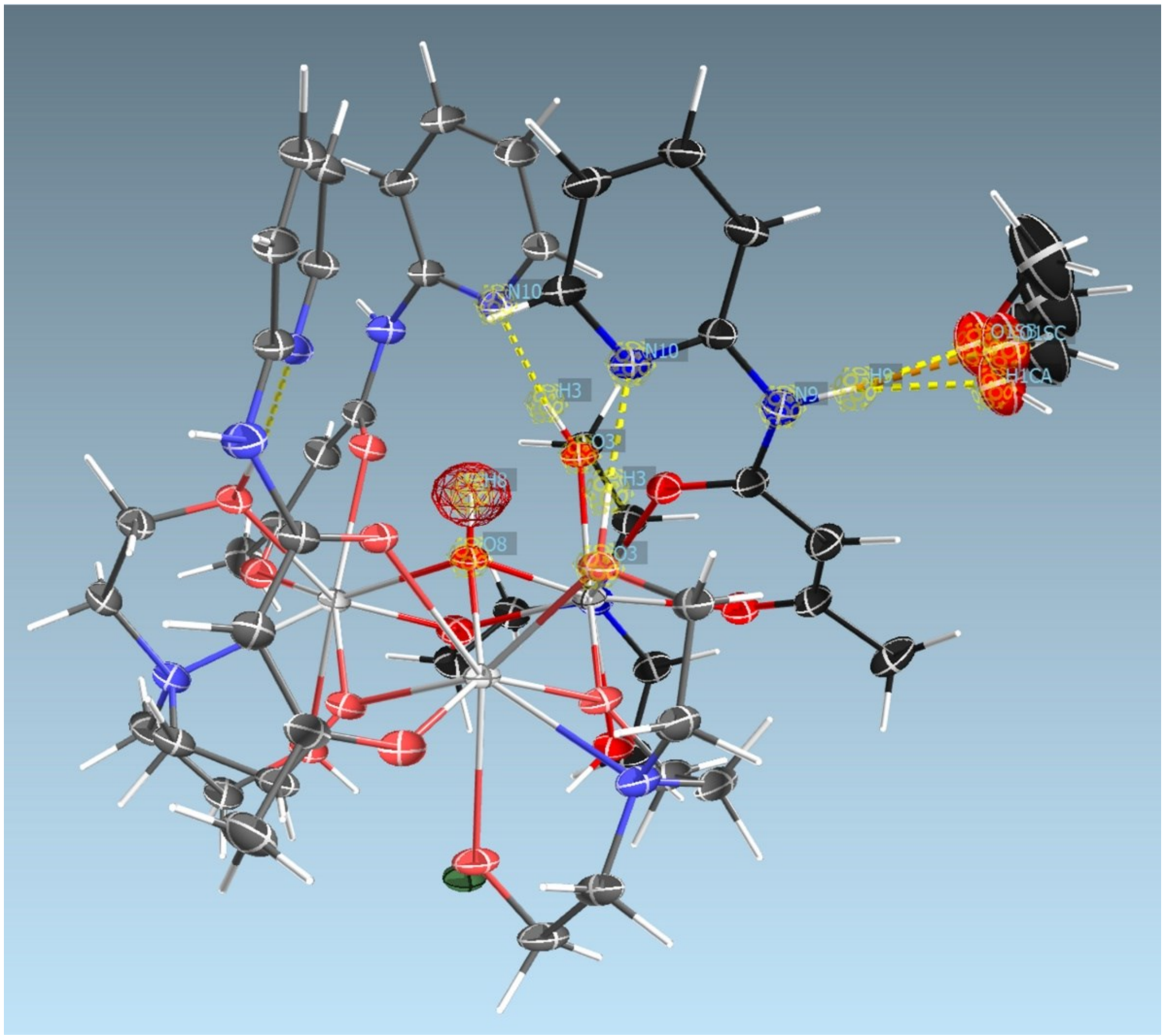

**Fig. S4 | Difference Fourier map of negative nuclear scattering density (red mesh) corresponding to the omitted H atom (H8) on the $\mu_3$-OH ligand if the neutron structure were solved as $[Dy_3(O)(teaH_2)_3(paa)_3]Cl\cdot 4MeOH\cdot PhCN$ with a monocation $[Dy_3(O)(teaH_2)_3(paa)_3]^+$ containing a bridging oxide ($\mu_3$-$O^{2-}$) ligand.**

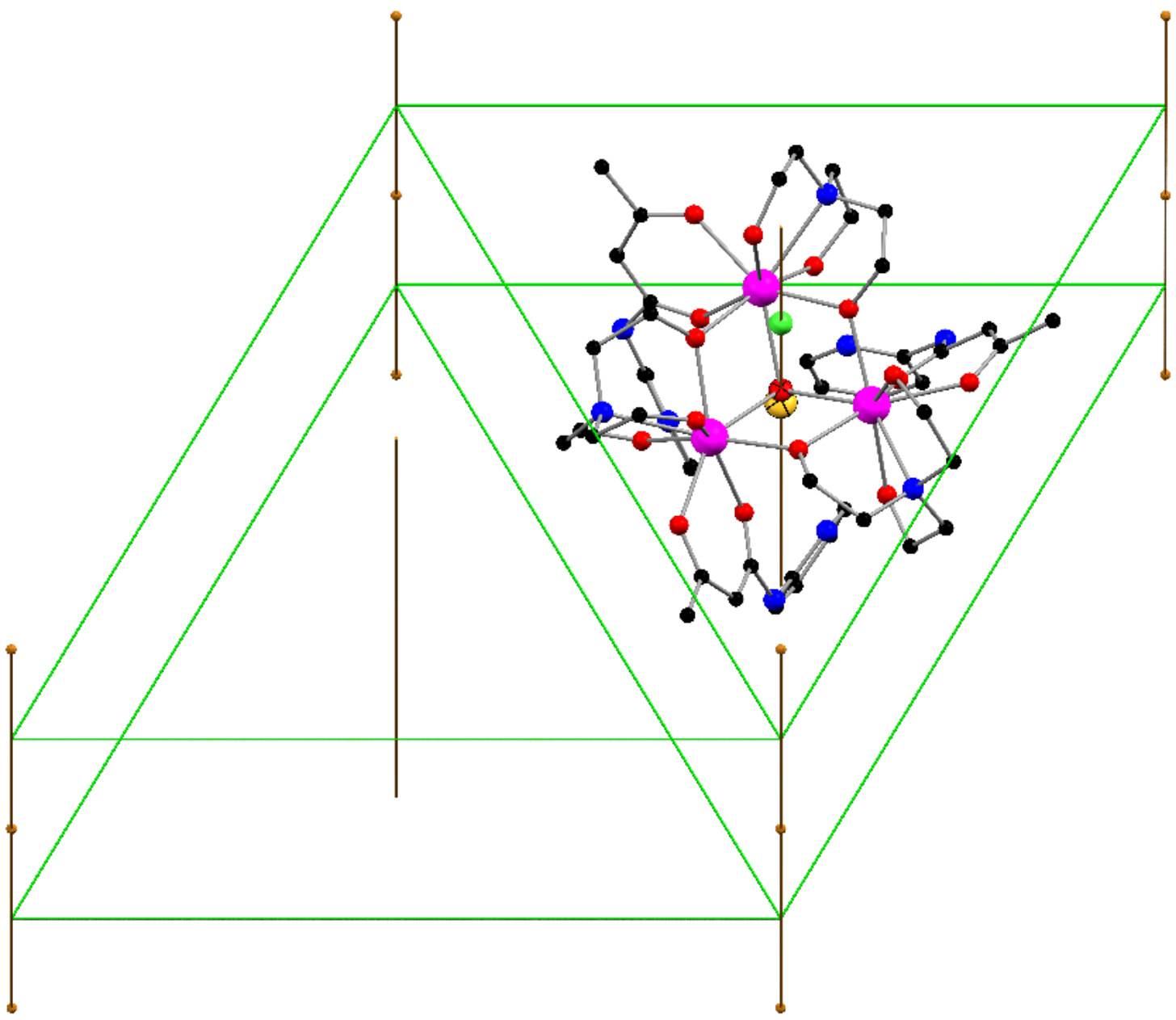

**Fig. S5 | Rotational symmetry elements in $Dy_3$-2.** Green and yellow lines are $C_2$ and $C_3$ axes, respectively. One $C_3$ axis is through the center of the **$Dy_3$-2** molecule.

## S2.2. ***Continuous shape measurements*** **(*CShMs*) for $Dy_3$-2 and $Dy_3$-1**

*CShM* was conducted to find the coordination geometry for $Dy^{3+}$ ions in both **$Dy_3$-2** and **$Dy_3$-1**. Cartesian coordinates were generated in the *CrystalExplorer* program[S10,11] using the structure of **$Dy_3$-2** by single-crystal X-ray diffraction, which were then used to make an input file for *Continuous Shape Measurement* (*CShM*) following typical procedures in the manual.[S12,13] Cartesian coordinates for **$Dy_3$-1** were taken from the previously reported structure.[S9] The $Dy^{3+}$ ion and N and O atoms of ligands bound to the metal atom in **$Dy_3$-2** are shown in Fig. S6. Results of the *CShM* results are given in Table S3. The *CShM* results (Table S3) show that the best fit for both **$Dy_3$-1** and **$Dy_3$-2** is the triangular dodecahedron.

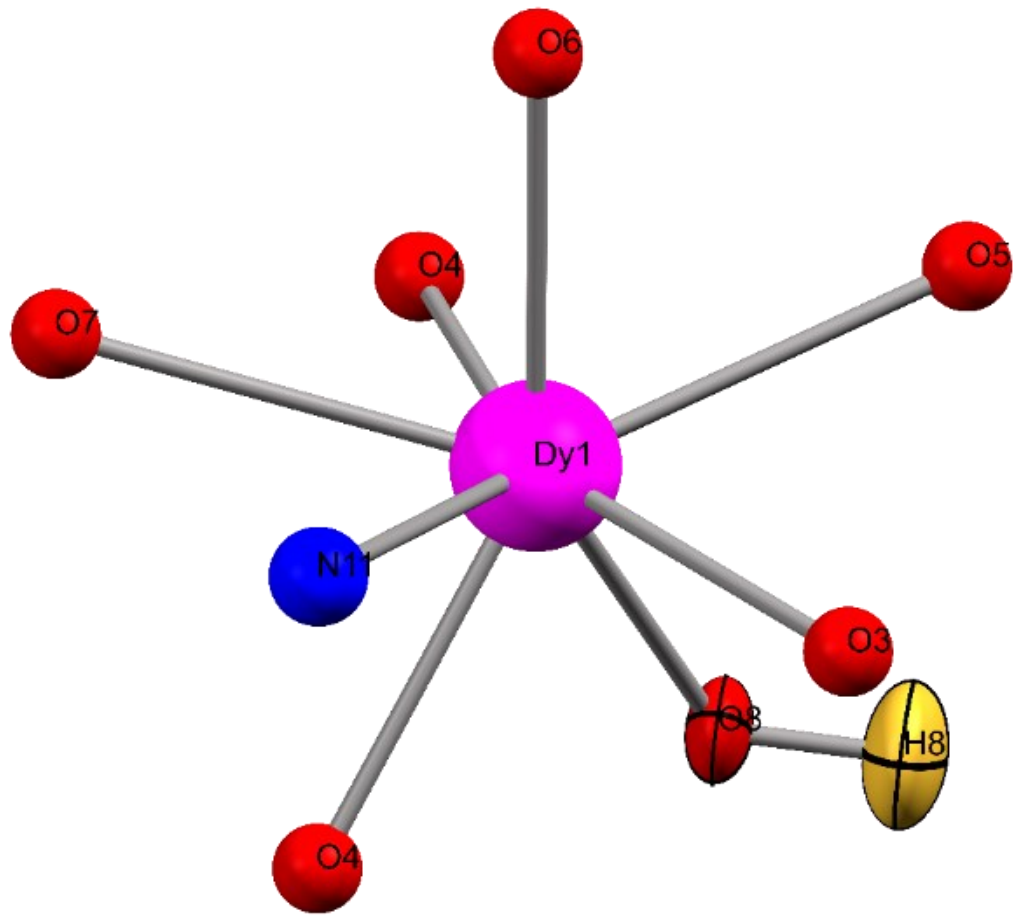

**Fig. S6 | Coordination of the $Dy^{3+}$ ion in $Dy_3$-2 from SCND.** The structure was used for *CShM*.

**Table S3**. *Continuous shape measurements* (*CShM*) for **$Dy_3$-2** and **$Dy_3$-1**. Values closer to zero show a better fit to the symmetry described.

| ***CShM*** | **Symmetry** | **$Dy_3$-2** | **$Dy_3$-1** |
|---|---|---|---|
| Octagon | $D_{8h}$ | 31.829 | 31.968 |
| Heptagonal pyramid | $C_{7v}$ | 23.728 | 23.736 |
| Hexagonal bipyramid | $D_{6h}$ | 14.289 | 14.214 |
| Cube | $O_h$ | 12.171 | 12.177 |
| Square antiprism | $D_{4d}$ | 2.508 | 2.579 |
| Triangular dodecahedron | $D_{2d}$ | 1.344 | 1.368 |
| Johnson - Gyrobifastigium (J26) | $D_{2d}$ | 9.743 | 9.686 |
| Johnson - Elongated triangular bipyramid (J14) | $D_{3h}$ | 26.979 | 26.938 |
| Johnson - Biaugmented trigonal prism (J50) | $C_{2v}$ | 2.150 | 2.142 |
| Biaugmented trigonal prism | $C_{2v}$ | 2.098 | 2.099 |
| Snub disphenoid (J84) | $D_{2d}$ | 2.378 | 2.378 |
| Triakis tetrahedron | $T_d$ | 12.591 | 12.665 |
| Elongated trigonal bipyramid | $D_{3h}$ | 22.562 | 22.551 |

### S2.3. Magnetic susceptibility measurements

DC Magnetic Susceptibility measurements were completed using a Quantum Design MPMS-7 T. Magnetization by variable temperature and variable magnetic field was collected. Measurements were taken of a powdered sample as well as single crystals oriented in multiple directions. Single crystals were coated with a thin layer of silicon grease to protect the crystal from degradation.

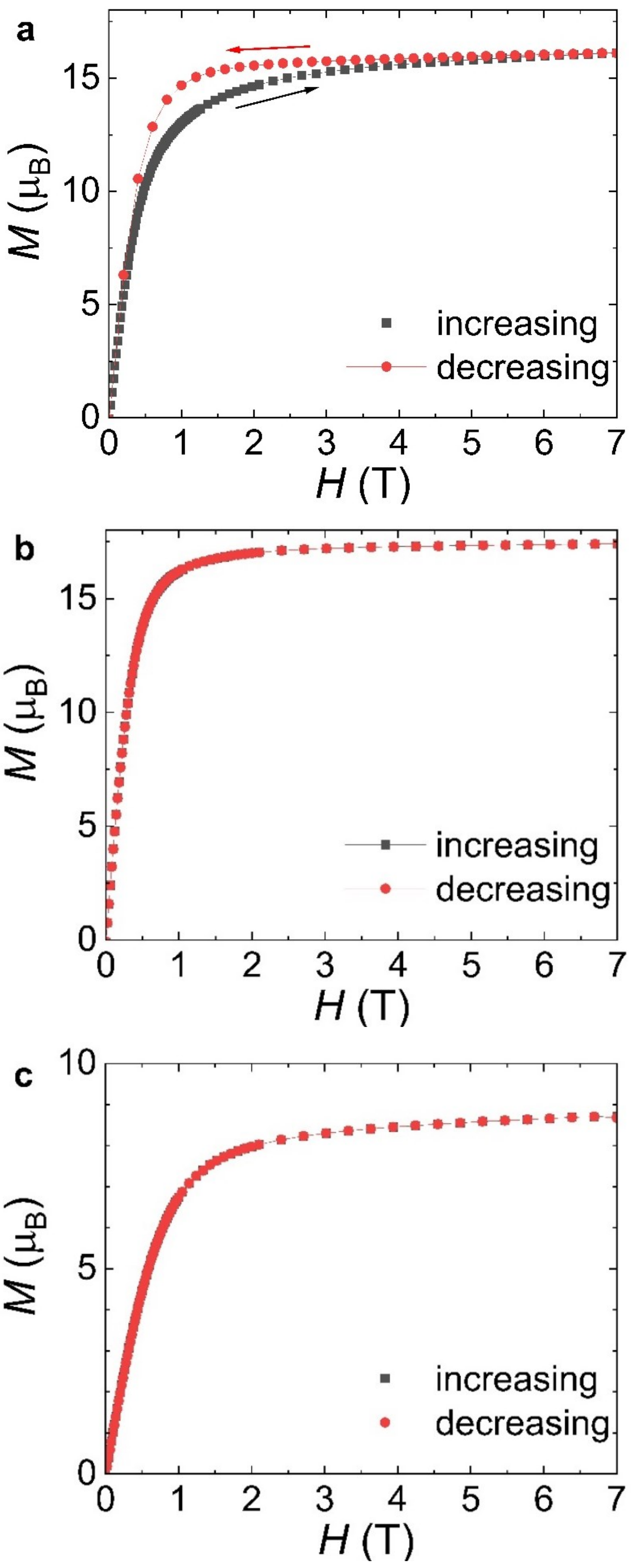

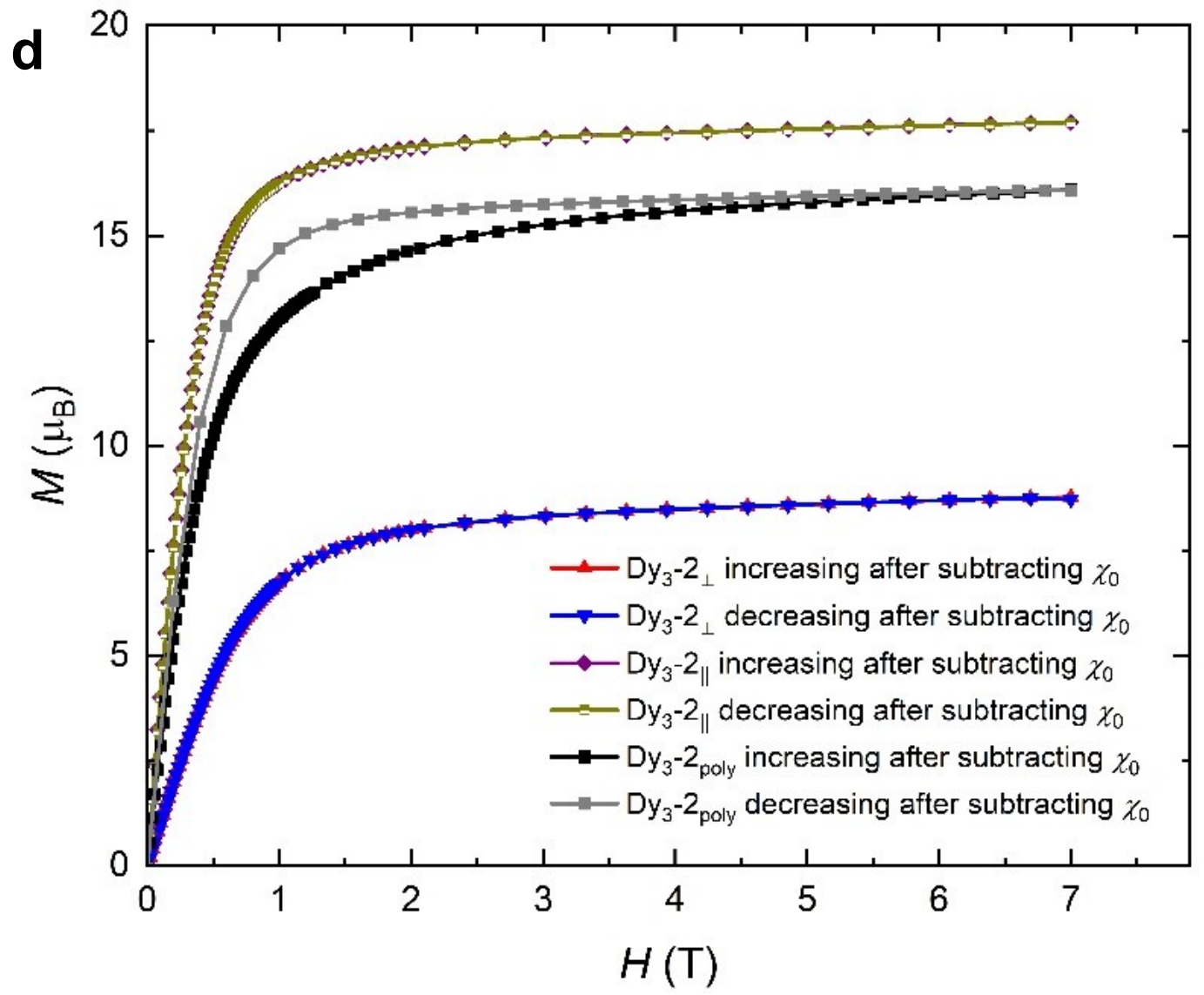

**Fig. S7 | Magnetization vs field at 1.9 K. a. $Dy_3$-2$_{poly}$. b. $Dy_3$-2$_{\parallel}$. c. $Dy_3$-2$_{\perp}$. d.** Data from **a-c** compiled together to show magnetic anisotropy of the single-crystal sample orientation.

## S.3. Computational details and additional results

The 4*f* electrons in lanthanide ions are strongly localized at the metal centers, leading to magnetic behavior dominated by crystal-field effects and strong spin–orbit coupling, while direct magnetic exchange interactions are comparatively weak. As a result, the electronic structure and magnetic properties of polynuclear lanthanide complexes are primarily dictated by the interplay between ligand-field splitting and spin–orbit coupling. A reliable description of these competing effects, therefore, necessitates the use of correlated post–Hartree–Fock *ab initio* methodologies.

Accordingly, the magnetic properties of the present polynuclear lanthanide system were investigated using a fragment-based approach, in which *ab initio* calculations were performed on individual lanthanide ions, followed by a model treatment of the magnetic interactions between the resulting single-ion fragments. The geometries of the models **1** and **2** were extracted from the crystal structure of **$Dy_3$-2**, with solvent molecules removed to slightly simplify the models. Initial guess orbitals for the metal centers were generated using the SEWARD module and subsequently employed in state-averaged complete active space self-consistent field (SA-CASSCF) calculations to obtain spin-free wavefunctions and energies. The active space was defined according to the 4*f*-electron configuration, with 9 electrons distributed over 7 4*f* orbitals (CAS (9,7)). Spin–orbit coupling effects were then incorporated via the RASSI-SO method, which constructs spin-orbit-coupled multiconfigurational states using the CASSCF wavefunctions as a basis.

For model **1**, the $Dy^{3+}$ ion has a $4f^9$ electronic configuration and a $^6H_{15/2}$ ground-state multiplet. Twenty-one sextet states were included in both the SA-CASSCF and RASSI-SO steps, a protocol that has been shown to yield reliable *g*-tensor estimates for $Dy^{3+}$ ions. The resulting spin–orbit-coupled multiplet energies and wavefunctions were subsequently analyzed using the SINGLE_ANISO module to derive the magnetic anisotropy parameters and *g*-tensors of the lowest-lying states (Tables S1–S3).

Density Functional Theory (DFT) calculations were performed using the B3LYP functional in the Gaussian 16 suite of programs to compute the exchange interaction constants in the isostructural $Gd_3$ complex (with all $Dy^{3+}$ ions in model **1** replaced by $Gd^{3+}$). These calculations are expected to provide an initial guess/order-of-magnitude for the underlying exchange interaction in the $Dy_3$ model **1**. Here, we have used, a double-zeta-quality basis set employing the Cundari–Stevens (CS) relativistic effective core potential for the Gd(III) atom, and the TZV basis set for the remaining atoms. The DFT calculations combined with the Broken Symmetry (BS) approach have been used to compute the magnetic exchange $J_{ex,Gd-Gd}$ value. We have employed the following Hamiltonian (S1), where $S_{Gd,1}$ and $S_{Gd,2}$ are the spins of the Gd(III) centers ($Sz$ = 7/2). The computed isotropic magnetic exchange interactions were $J_{ex,Gd-Gd}$ = -0.13 $cm^{-1}$ (inter-triangular), which were rescaled by multiplying by $5^2/7^2$ to convert to the Lines model for Dy(III) ($J_{ex,\mathrm{Dy-Dy}}$ = –0.06 $cm^{-1}$). These were the initial guesses for the isotropic exchange between Dy-Dy ions in the $Dy_3$ complex. The exchange interactions were found to be antiferromagnetic and very weak in magnitude (only the exchange part), which is typical for lanthanide-based systems. The Hamiltonian used for the DFT exchange calculations is as follows (Eq. S1):

$$H_{Gd-Gd} = -2J_{ex,Gd-Gd}\, S_{Gd,1}\, S_{Gd,2} \qquad \text{(S1)}$$

The effective Heisenberg Hamiltonian utilized in the Lines model is given in the below equation (Eq. S2):

$$H = -\sum_{\substack{i,j=1 \\ i\neq j}}^{3} J_{ex} S_i \cdot S_j \qquad \text{(S2)}$$

in which $S_i$ and $S_j$ correspond to local spin operators ($S$ = 5/2) on the $i^{th}$ and $j^{th}$ sites, respectively.

The dipolar contribution to the magnetic exchange coupling can be calculated by the following equation (Eq. S3):

$$J_{\text{Ising,(dip)}} = \{\mu_0/4\pi\}(\mu_B g_{zi}\, g_{zj})(\cos\theta_{ij} - 3\cos\theta_{in}\cos\theta_{jn}/r^3) \quad \text{(S3)}$$

Here $g_{zi,j}$ are the z-component of the $g$ tensor within pseudo $S_{iz}$ = 1/2 formalism, $\mu_B$ is the Bohr magneton, $r$ is the distance between interacting ions, $\theta_{ij}$ is the angle between the easy axes of the interacting ions, and $\theta_{in,jn}$ are the angles between the corresponding easy axis and the vector connecting two interacting ions.

$J_{ex}$ was converted to $J_{\text{Ising, (ex)}}$ exchange parameter using the following equation (Eq. S4):

$$J_{\text{Ising, (ex)}} = 25 J_{ex} \cos\delta_{ij} \quad \text{(S4)}$$

in which $J_{ex}$ and $J_{\text{Ising, (ex)}}$ are the exchange coupling constant in noncollinear Ising and Lines models, respectively, and $\delta_{ij}$ is the angle between the main magnetic axes of the two interacting sites $i$ and $j$.

The exchange/dipolar interactions between neighboring $Dy^{III}$–$Dy^{III}$ ions in model **1** have been computed by fitting with the experimental magnetic data using the Lines model[S14] within the POLY_ANISO routine. Table S4 gives a summary of the results with details for each $Dy^{3+}$ ion listed in Tables S5-S7.

Despite the highly axial coordination environment,[S15 S16] the overall splitting of the $^6H_{15/2}$ multiplet remains relatively small (<540 cm$^{-1}$), with a modest stabilization of the ground KD (~112 cm$^{-1}$). For each $Dy^{3+}$ ion, the angle between the principal $g_{zz}$ axes of the ground and first excited KDs is approximately 120°, indicating that magnetic relaxation is expected to proceed predominantly through the first excited KD at all three equivalent $Dy^{3+}$ centers. The corresponding calculated energy barrier is $U_{cal}$ = 112 cm$^{-1}$.

**Table S4**. Summary of CASSCF/RASSI-SO SINGLE_ANISO computed energies of eight KDs of model **1**, ground state computed *g*-tensor and angle of anisotropy axis with $Dy_3$ plane.

| Kramers Doublets | **Dy-1** | **Dy-2** | **Dy-3** |
|---|---|---|---|
| 1 | 0.0 | 0.0 | 0.0 |
| 2 | 112.1 | 112.5 | 111.8 |
| 3 | 176.0 | 175.8 | 171.0 |
| 4 | 263.3 | 263.6 | 255.3 |
| 5 | 343.8 | 344.4 | 333.1 |
| 6 | 388.5 | 389.5 | 374.6 |
| 7 | 483.2 | 483.4 | 467.6 |
| 8 | 537.5 | 535.8 | 518.4 |
| Main values of *g* tensors | | | |
| $g_{xx}$ | 0.036 | 0.036 | 0.037 |
| $g_{yy}$ | 0.049 | 0.049 | 0.050 |
| $g_{zz}$ | 19.687 | 19.719 | 19.697 |
| Angle of anisotropy axis with $Dy_3$ plane (θ) | 17.37° | 17.31° | 17.72° |
| Canting angle (φ) | 38.36° | 40.11° | 40.96° |

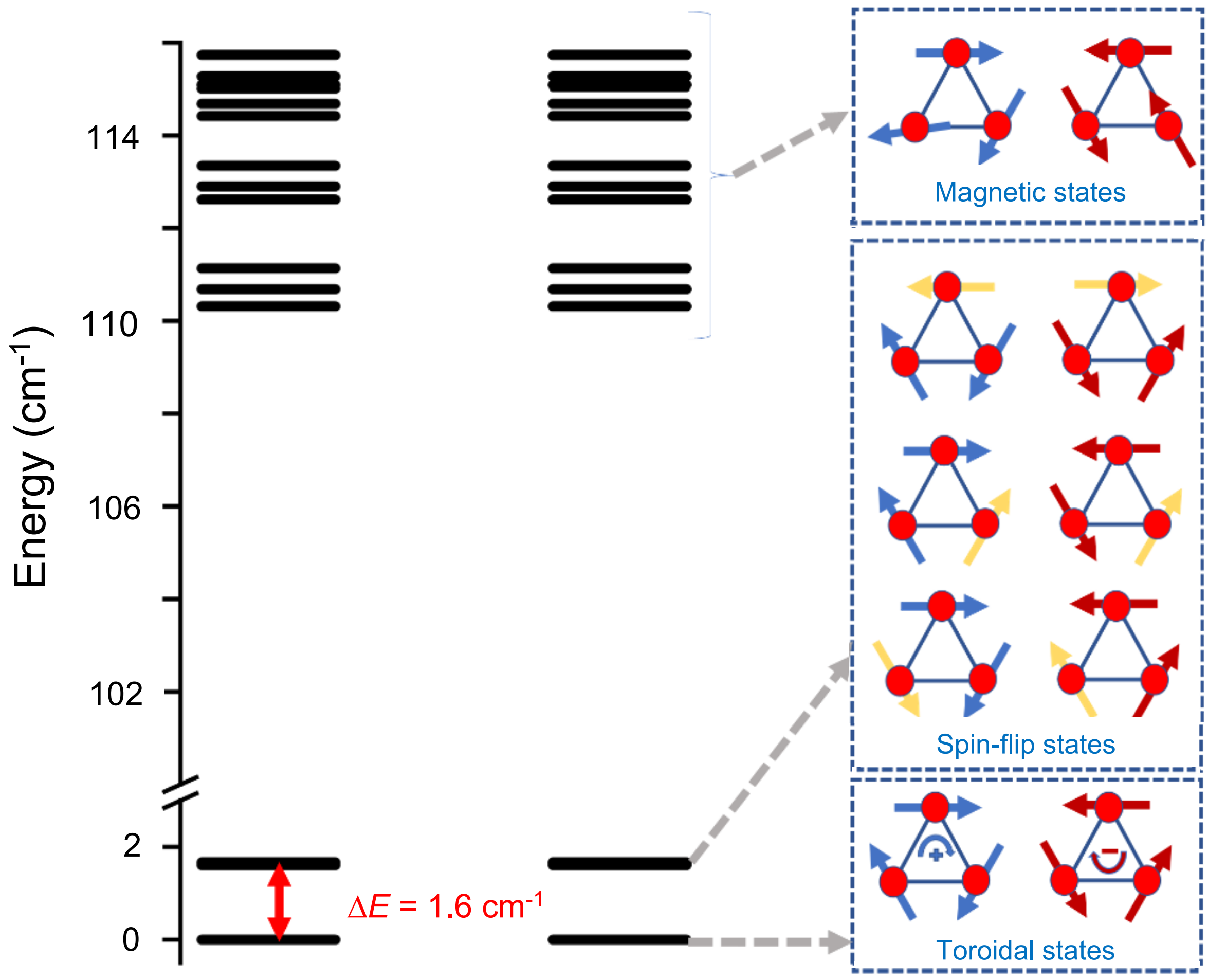

**Fig. S8 | Additional results of *ab initio* calculations.** Low-energy spectrum of model **1**.

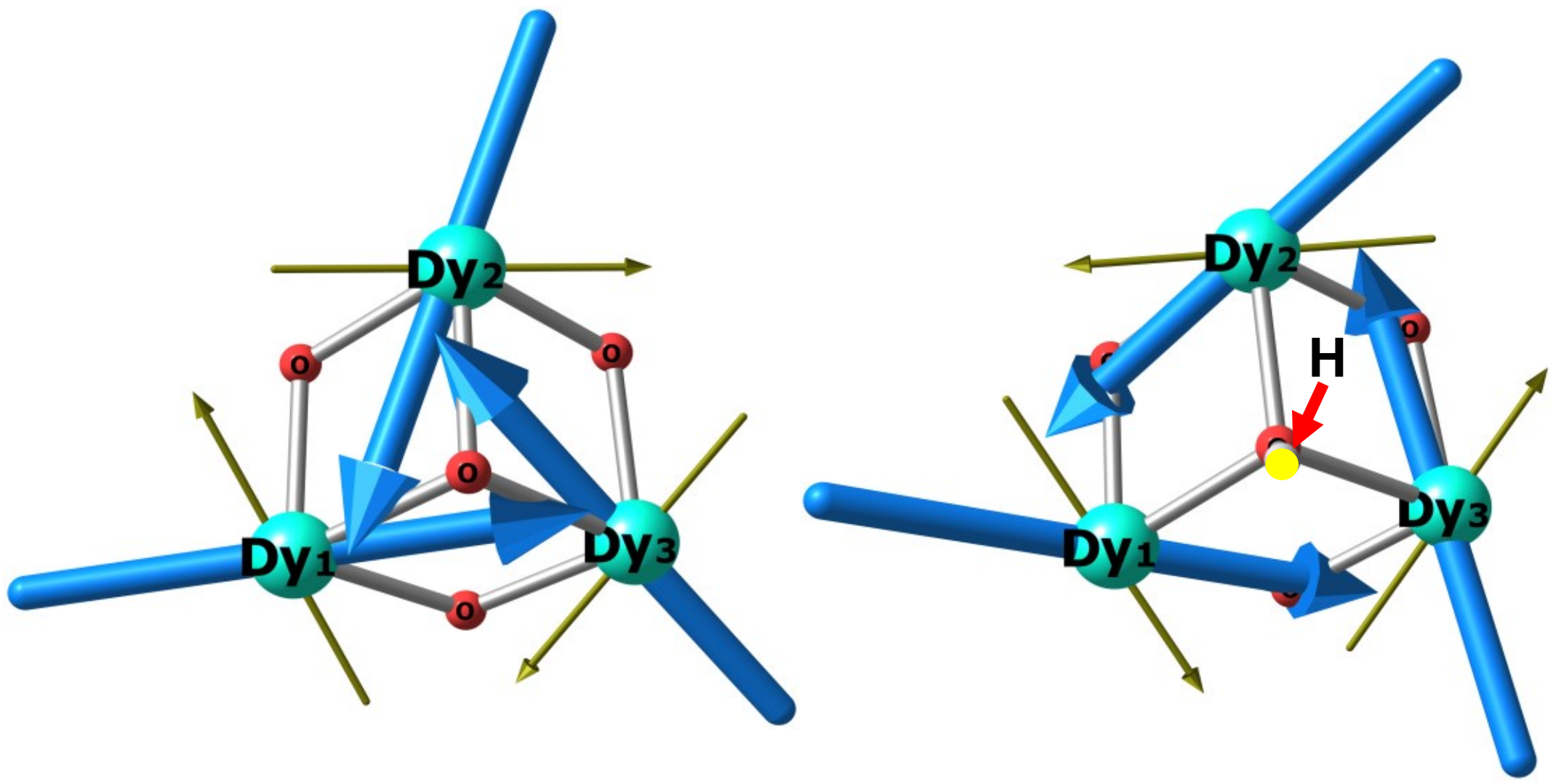

**Fig. S9 | Calculated anisotropy directions**. **a.** Model **2** with a bridging $O^{2-}$ (oxide) ligand. **b**. Model **1** (for **$Dy_3$-2**) with a bridging $OH^-$ ligand. Only the $Dy_3$ cores are shown for clarity. The yellow circle represents the H atom in the bridging $OH^-$ ligand. The yellow-green arrows are the ideal arrows for the net-toroidal moment, exactly at the plane of the Dy triangle. The canting angle (φ) and out-of-plane angle (θ) are calculated with respect to these arrows.

**Table S5**. CASSCF+RASSI-SO+SINGLE_ANISO calculated energy spectrum of the eight lowest lying Kramers doublets, *g* tensors and angles (θ) of the principal anisotropy axes of excited states (ES) with respect to the ground state (GS) for Dy-1 site in model **1**.

| Kramers doublets | Energy ($cm^{-1}$) | $g_x$ | $g_y$ | $g_z$ | Angle (°) | Wavefunctions |
|---|---|---|---|---|---|---|
| 1 | 0 | 0.036 | 0.049 | 19.687 | | 98%\|±15/2> |
| 2 | 112.1 | 0.455 | 0.591 | 17.578 | 124.69 | 53%\|±13/2>+38%\|±1/2> |
| 3 | 176.0 | 0.575 | 1.446 | 13.692 | 33.00 | 76%\|±13/2>+34%\|±1/2> |
| 4 | 263.3 | 3.947 | 5.890 | 10.512 | 140.49 | 66%\|±11/2>+42%\|±9/2> |
| 5 | 343.8 | 10.048 | 6.193 | 0.143 | 44.68 | 52%\|±13/2>+38%\|±1/2> |
| 6 | 388.5 | 2.293 | 3.824 | 14.530 | 94.31 | 58%\|±5/2>+55%\|±3/2> |
| 7 | 483.2 | 0.067 | 0.403 | 17.510 | 57.49 | 57%\|±9/2>+49%\|±7/2> |
| 8 | 537.5 | 0.090 | 0.282 | 19.049 | 88.33 | 52%\|±1/2>+48%\|±3/2> |

**Table S6**. CASSCF+RASSI-SO+SINGLE_ANISO calculated energy spectrum of the eight lowest lying Kramers doublets, *g* tensors and angles (θ) of the principal anisotropy axes of excited states (ES) with respect to the ground state (GS) for Dy-2 site in model **1**.

| Kramers doublets | Energy ($cm^{-1}$) | $g_x$ | $g_y$ | $g_z$ | Angle (°) | Wavefunctions |
|---|---|---|---|---|---|---|
| 1 | 0 | 0.036 | 0.049 | 19.719 | | 98%\|±15/2> |
| 2 | 112.5 | 0.447 | 0.585 | 17.601 | 124.67 | 52%\|±13/2>+38%\|±1/2> |
| 3 | 175.8 | 0.578 | 1.442 | 13.716 | 32.99 | 76%\|±13/2>+34%\|±3/2> |
| 4 | 263.6 | 3.983 | 5.900 | 10.498 | 140.71 | 65%\|±11/2>+42%\|±9/2> |
| 5 | 344.4 | 10.151 | 6.148 | 0.061 | 44.82 | 60%\|±7/2>+44%\|±11/2> |
| 6 | 389.5 | 2.243 | 3.724 | 14.703 | 94.44 | 58%\|±5/2>+42%\|±1/2> |
| 7 | 483.4 | 0.062 | 0.395 | 17.553 | 57.36 | 57%\|±9/2>+49%\|±7/2> |
| 8 | 535.8 | 0.085 | 0.275 | 19.093 | 88.65 | 52%\|±1/2>+48%\|±3/2> |

**Table S7**. CASSCF+RASSI-SO+SINGLE_ANISO calculated energy spectrum of the eight lowest lying Kramers doublets, *g* tensors and angles (θ) of the principal anisotropy axes of excited states (ES) with respect to the ground state (GS) for Dy-3 site in model **1**.

| Kramers doublets | Energy ($cm^{-1}$) | $g_x$ | $g_y$ | $g_z$ | Angle (°) | Wavefunctions |
|---|---|---|---|---|---|---|
| 1 | 0 | 0.037 | 0.050 | 19.697 | | 98%\|±15/2> |
| 2 | 111.8 | 0.540 | 0.730 | 17.644 | 121.20 | 49%\|±13/2>+33%\|±11/2>+ 31%\|±9/2> |
| 3 | 171.0 | 0.537 | 1.617 | 13.631 | 31.51 | 77%\|±13/2>+34%\|±11/2> |
| 4 | 255.3 | 4.201 | 6.317 | 10.227 | 139.6 | 64%\|±13/2>+33%\|±1/2> |
| 5 | 333.1 | 10.609 | 5.844 | 0.235 | 42.04 | 60%\|±7/2>+36%\|±5/2> |
| 6 | 374.6 | 2.117 | 3.635 | 14.710 | 95.17 | 55%\|±3/2>+42%\|±1/2> |
| 7 | 467.6 | 0.017 | 0.361 | 17.543 | 56.66 | 57%\|±9/2>+47%\|±7/2> |
| 8 | 518.4 | 0.091 | 0.299 | 19.036 | 88.96 | 51%\|±1/2>+48%\|±3/2> |

**Table S8.** Crystal field parameters for each Dy(III) center for model **1**. The quantization axis is chosen to be the main magnetic axis of the ground pseudo-doublet.

| ***k*** | ***q*** | $B_k^q$ | $B_k^q$ | $B_k^q$ |
|---|---|---|---|---|
| | | **Dy-1** | **Dy-2** | **Dy-2** |
| 2 | -2 | **4.71E-01** | **-4.08E+00** | **5.91E-02** |
| | -1 | **-9.03E+00** | **-3.68E+00** | **-1.15E+00** |
| | 0 | **-4.38E+00** | **-4.35E+00** | **-4.04E+00** |
| | 1 | **-1.14E+00** | **-8.42E+00** | **8.70E+00** |
| | 2 | **4.46E+00** | **-1.86E+00** | **-4.33E+00** |
| 4 | -4 | 1.26E-01 | -6.66E-02 | 1.28E-01 |
| | -3 | 3.86E-01 | -3.93E-01 | -1.01E-01 |
| | -2 | 9.91E-02 | 6.43E-02 | -1.15E-01 |
| | -1 | 7.06E-02 | 3.62E-02 | -1.15E-02 |
| | 0 | -3.44E-02 | -3.42E-02 | -3.50E-02 |
| | 1 | -5.63E-03 | 5.29E-02 | -3.63E-02 |
| | 2 | -1.32E-01 | 1.55E-01 | 1.29E-01 |
| | 3 | 8.93E-02 | -9.22E-02 | 3.50E-01 |
| | 4 | -5.74E-03 | -1.04E-01 | 1.74E-02 |
| 6 | -6 | 5.53E-03 | 5.25E-03 | -3.17E-03 |
| | -5 | -4.31E-03 | -1.53E-02 | -1.69E-02 |
| | -4 | -7.17E-04 | 1.25E-03 | -1.25E-03 |
| | -3 | 3.15E-03 | -3.22E-03 | 1.23E-03 |
| | -2 | 5.09E-04 | -9.78E-04 | -3.26E-04 |
| | -1 | 3.76E-03 | 1.96E-03 | 1.08E-04 |
| | 0 | 8.11E-05 | 7.21E-05 | 2.35E-05 |
| | 1 | -7.52E-05 | 3.17E-03 | -3.87E-03 |
| | 2 | 9.75E-04 | 4.20E-05 | -5.20E-04 |
| | 3 | -2.00E-03 | 1.88E-03 | 3.84E-03 |
| | 4 | 9.39E-04 | 2.11E-04 | 6.49E-04 |
| | 5 | -1.54E-02 | -5.05E-03 | -1.37E-03 |
| | 6 | 5.60E-03 | 5.70E-03 | -6.90E-03 |
| 8 | -8 | -8.61E-06 | 1.88E-06 | -7.54E-06 |
| | -7 | 1.51E-05 | -3.80E-05 | -5.84E-05 |
| | -6 | 6.33E-06 | 6.46E-06 | -3.51E-06 |
| | -5 | -4.95E-06 | 3.49E-05 | 1.76E-05 |
| | -4 | 4.39E-06 | -1.91E-06 | 5.42E-06 |
| | -3 | 1.58E-05 | -2.74E-05 | 6.88E-06 |
| | -2 | 5.75E-06 | 1.71E-06 | -9.29E-06 |
| | -1 | 2.06E-06 | 1.36E-06 | -5.10E-07 |
| | 0 | -1.64E-07 | -1.69E-07 | -1.91E-07 |
| | 1 | 1.58E-07 | 1.34E-06 | -1.42E-06 |
| | 2 | -6.86E-06 | 1.20E-05 | 8.08E-06 |
| | 3 | -3.59E-06 | 3.81E-06 | 2.30E-05 |
| | 4 | -9.26E-08 | -4.82E-06 | 9.51E-07 |
| | 5 | 1.35E-05 | -1.47E-05 | -2.39E-05 |
| | 6 | 1.18E-05 | 1.04E-05 | -1.25E-05 |
| | 7 | 4.84E-05 | 4.60E-05 | 2.46E-06 |
| | 8 | 2.95E-06 | -8.55E-06 | -2.95E-07 |
| 10 | -10 | 7.09E-07 | -8.79E-07 | -9.97E-07 |
| | -9 | -2.99E-07 | -5.10E-07 | 3.32E-07 |
| | -8 | -4.51E-07 | 5.91E-07 | -2.38E-07 |
| | -7 | -6.61E-07 | -1.95E-06 | -1.57E-06 |
| | -6 | 7.19E-09 | 5.25E-08 | -1.16E-07 |
| | -5 | -1.84E-07 | 1.68E-08 | 4.49E-08 |
| | -4 | 2.43E-07 | -5.50E-07 | 3.51E-07 |
| | -3 | -5.10E-07 | 6.50E-07 | 2.27E-07 |
| | -2 | -1.38E-09 | 3.64E-08 | -1.44E-08 |
| | -1 | -9.26E-08 | -4.30E-08 | -3.44E-09 |
| | 0 | -2.07E-09 | -2.28E-09 | -1.21E-09 |
| | 1 | -1.07E-09 | -8.07E-08 | 1.05E-07 |
| | 2 | -4.83E-08 | 2.82E-08 | 3.07E-08 |
| | 3 | -1.82E-07 | 1.24E-07 | -6.48E-07 |
| | 4 | -4.58E-07 | 8.57E-08 | -3.94E-07 |
| | 5 | 1.54E-07 | 2.19E-07 | 2.60E-07 |
| | 6 | -1.81E-07 | -1.89E-07 | 1.39E-07 |
| | 7 | 1.30E-06 | 3.04E-07 | -1.58E-06 |
| | 8 | -3.84E-07 | -1.47E-07 | -4.77E-07 |
| | 9 | 6.57E-07 | -2.62E-07 | 6.35E-07 |
| | 10 | -6.20E-07 | -3.37E-07 | 8.70E-08 |

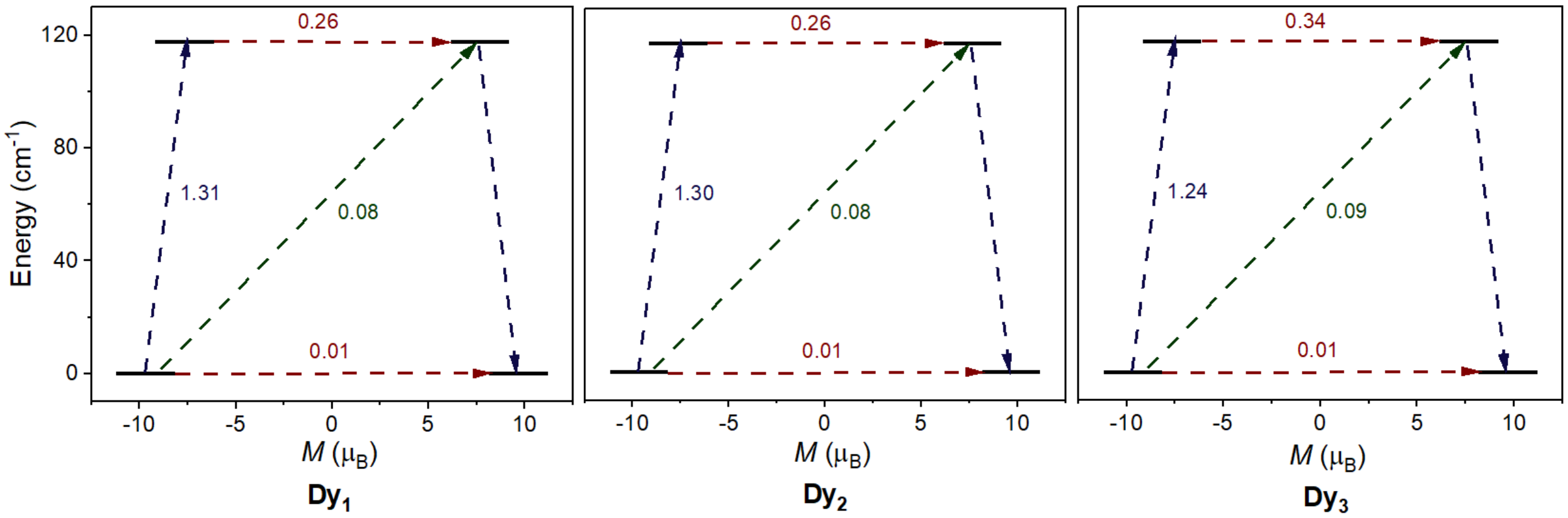

**Fig. S10 | Computed blocking barrier for the Dy1–Dy3 sites in the $Dy_3$ unit in model 1.** The thick black lines represent the Kramers doublets plotted as a function of the calculated magnetic moment. The blue and green arrows indicate possible relaxation pathways via Orbach and Raman processes, respectively. The dotted red lines denote the presence of QTM and thermally assisted QTM (TA-QTM) between the connected doublets. The numbers associated with each arrow correspond to the mean absolute values of the transition magnetic moment matrix elements.

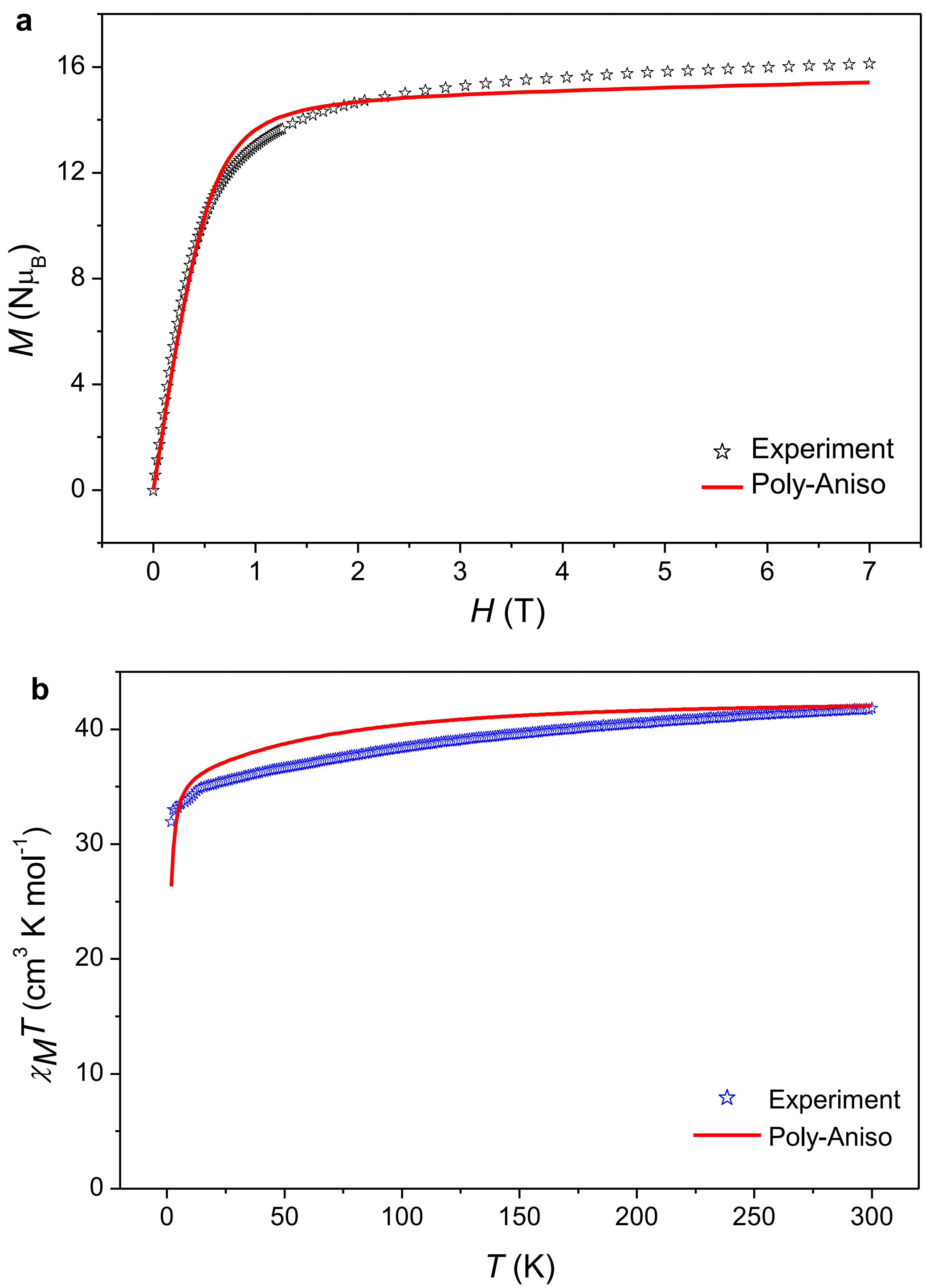

**Fig. S11 | Comparison of experimental and calculated results. a.** Experimental powder magnetization (*M* vs *H*) data (symbols) with the corresponding POLY_ANISO fit (red line). **b.** Experimental magnetic susceptibility ($\chi_M T$ vs $T$) data (symbols) together with the POLY_ANISO fit (red line).

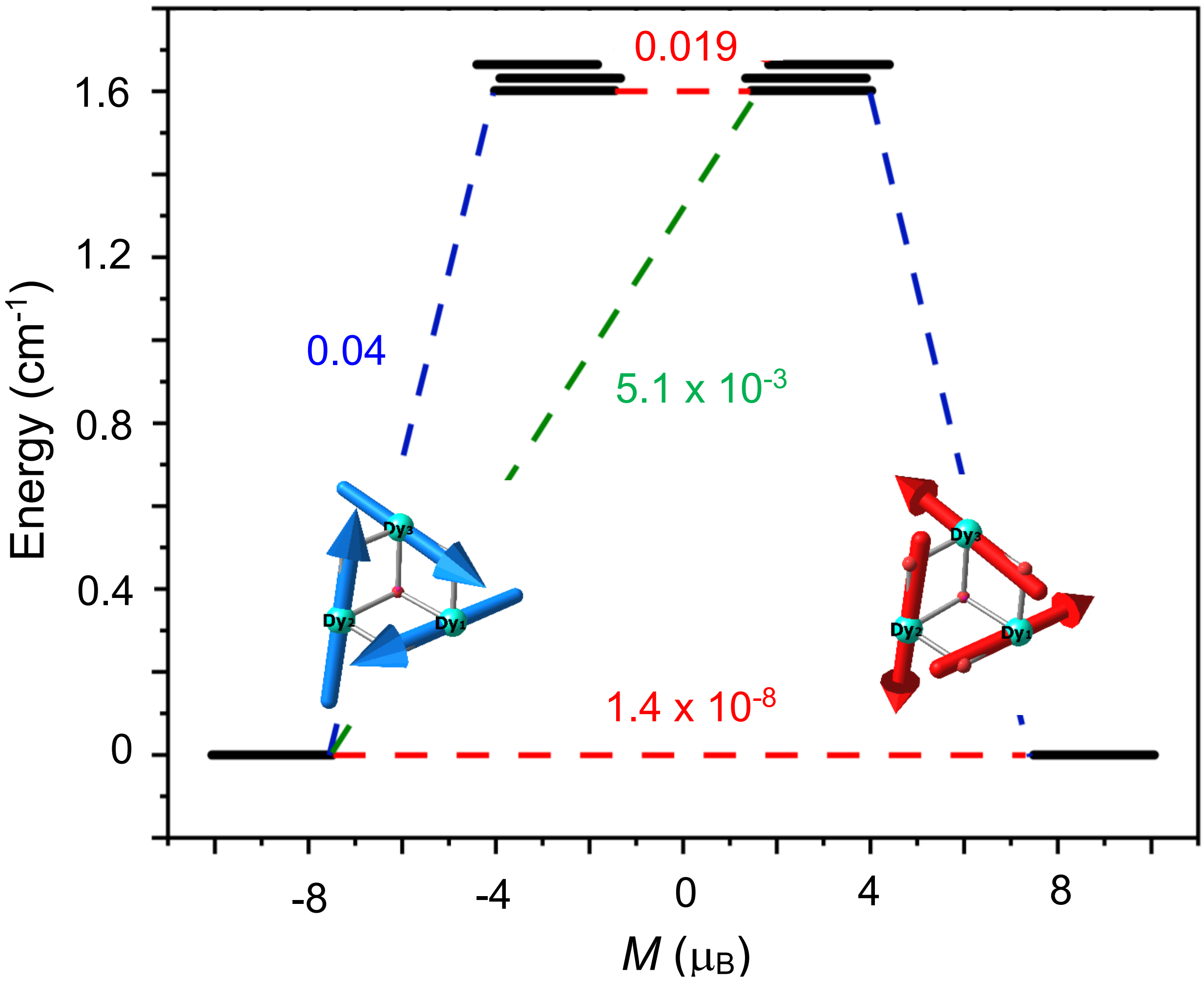

**Fig S12 | Low-lying exchange energies in the $Dy_3$ core in model 1.** Solvents and H atoms are removed for clarity. Every exchange state (represented by thick black lines) has been arranged based on the corresponding magnetic moment. The red dotted lines indicate a tunnelling transition ($\Delta_{tun}$; tunnel splitting or tunnel gaps) within each doublet. The blue and green dotted lines indicate the transition-moment integrals for Orbach and Raman relaxation processes.

## S4. References in SI